# Towards Wafer-Scale Production of Two-Dimensional Transition Metal Chalcogenides

Peijian Wang,[†,*] Deren Yang,[†,§] Xiaodong Pi[†,§*]

[†] Hangzhou Global Scientific and Technological Innovation Center, Zhejiang University, Hangzhou, Zhejiang 311215, China

[§] State Key Laboratory of Silicon Materials and School of Materials Science and Engineering, Zhejiang University, Hangzhou, Zhejiang 310007, China

## Abstract

Two-dimensional (2D) Transition Metal Chalcogenides (TMCs) have attracted tremendous interest from both the scientific and technological communities due to their variety of properties and superior tunability through layer number, composition, and interface engineering. Wafer-scale production of 2D TMCs is critical to the industrial applications of these materials. Extensive efforts have been bestowed to the large-area growth of 2D TMCs through various approaches. In this review, recent advances in obtaining large-area 2D TMCs by different methods such as chemical vapor deposition (CVD), metal-organic CVD, physical vapor deposition are firstly highlighted and their advantages and disadvantages are also evaluated. Then strategies for the control of the grains, morphology, layer number and phase to achieve controllable and uniform thicknesses and large crystal domains for 2D TMCs are discussed. Applications of large-area 2D TMCs in electronics, optoelectronics, spintronics etc. are also introduced. Finally, ideas and prospects for the future developments of wafer-scale 2D TMCs are provided.

# 1. Introduction

Materials play a significant role in the development of human civilization. In the 1960s, people started entering the Age of Information Technology. Such a shift has been intensified since the 1990s. The intrinsic dilemma between the miniaturization of semiconductor devices based on the currently prevalent silicon (Si) material and the quantum interference induced by the extremely small size calls for brand-new micro-nano-scale materials.[1-6] Since the 1980s, the significant role of dimensionality has been recognized apart from the bulk phase, inspiring the swift development of low-dimensional materials (0D, 1D, 2D).[7-10] The realm of 0D quantum dots,[11-14] 1D nanowire,[15-19] nanotube[20-24] etc. has become hot research spots in succession. However, 2D materials remained to be a silent zone somehow. When the timeline reached 2004, graphene, the single layer of graphite, was discovered through a scotch-tape exfoliation method and was found to have ultra-high mobility.[25-27] Various 2D materials such as transition metal chalcogenides (TMCs),[28-30] boron nitrides (BN) [31, 32], black phosphorus[33, 34], and some new materials including Si,[35] Te,[36] and black arsenic-phosphorus[37] were subsequently synthesized, enabling the fabrication of numerous novel devices. The field of 2D materials is now making strong voice, which is no longer ignored by the scientific and technological community.[38-40]

Among the various 2D materials, the TMCs constituted by the transition metal elements (M) and the chalcogen elements (X) are a large family. M can range from group 4B to group 10 in the periodic table, while X is S, Se or Te.[41-47] Figure 1a exhibits the binary TMCs that have already been explored. This variation of composition of elements transfers to a wide range of properties ranging from insulators (e.g. $HfS_2$),[48, 49] semiconductors (e.g.

$MoS_2$, $WS_2$),[50, 51] semi metals (e.g. $MoTe_2$, $TiSe_2$),[52, 53] to metals (e.g. $NbSe_2$, $VS_2$)[54-56]. Furthermore, TMCs can form alloys in the formula such as $WSe_{2-x}S_x$, with composition-tunable properties.[57-59] Unlike semimetallic graphene, semiconducting TMCs usually possess a bandgap. For example, $MoS_2$ exhibits a bandgap, transitioning from indirect bandgap to direct bandgap when it thins from bulk phase to monolayer (Figure 1b), boosting the photoluminescence (PL) efficiency.[51, 60-62] Additionally, inversion symmetry

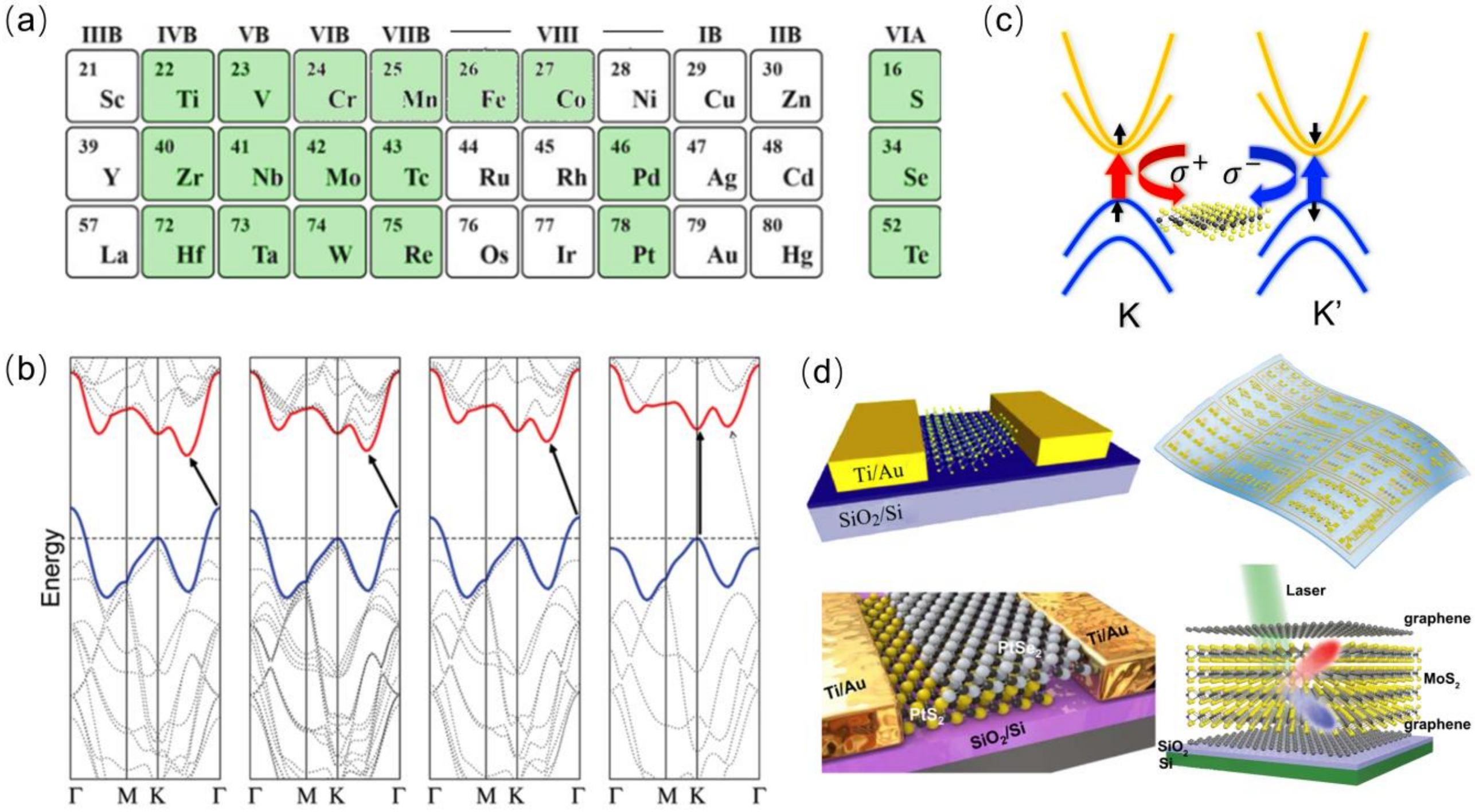

Figure 1. (a) The portion of the periodic table of the transition metal and chalcogen elements, with the green color standing for the 2D TMCs which have been explored in the research. (b) Transition of the band gap of $MoS_2$ from indirect bandgap to direct band gap. From left to right: bulk, quadrilayer, bilayer, and monolayer. (c) The two energetic degenerate but inequivalent valleys in K and K' points, giving out two polarized emissions. (d) Schematic diagram showing the conceptual electronic and optoelectronic devices fabricated from 2D TMCs. (b) Reproducedd with permission from ref 60. Copyright 2010, American Chemical Society. (d) Reproduced with permission from ref 83, 90, 89, 91. Copyright 2015, 2020, 2018, 2013. Springer Nature.

breaking and large spin-orbit interaction, resulting in two energetic degenerate but inequivalent valleys in the band structure of a TMC (Figure 1c), may lead to breakthroughs in valleytronics and spintronics.[63-70] A few TMCs (e.g. $NbSe_2$, $TaS_2$, $VSe_2$) have shown superconductivity,[71, 72] charge density wave[73, 74] and Mott transition (transition from metal to non-metal)[75, 76]. In addition, some novel TMCs such as $Fe_3GeTe_2$ display intrinsic magnetism with gate-tunable Curie points.[66, 77] Semi-metallic TMCs like $WTe_2$ show topological structures in the energy band, enabling the study on topological physics.[78, 79] The rich intrinsic properties and superior tunability of TMCs hold great promise for all kinds of technologically important applications in electronics, optoelectronics, spintronics, valleytronics etc. (Figure 1d).[51, 80-91]

Due to numerous unique properties of 2D TMCs, devices such as transistors, photodetectors and photodiodes based on 2D TMCs have been actively pursued. The ultimate destination of the miniaturization of the electronic devices will be one atomic thickness. 2D materials are the most adequate candidate. New-generation electronic and optoelectronic devices may be well built on 2D TMCs.[51, 92-95] For excellent scalability for large-scale integrated circuits and uniformity of the fabricated devices, wafer-scale production is the critical step towards the industrial applications of 2D TMCs.[84, 96] Among the 2D materials, the wafer-scale growth is most progressive for graphene. Researchers have used copper foil or Ge (110) to grow large-area graphene single crystal or annealed SiC to sublime Si to obtain graphene.[97-101] Recently, Wu et al. have reportedly fabricated large single-crystal copper foils with high-index facets and grown epitaxial graphene on these foils with the size of A4 paper.[102] People have also attempted to fabricate wafer-scale 2D TMCs and attained considerable achievements. The size of 2D TMCs has recently increased from several micrometers to a few decimeters.[103-105] Methods employed to fabricate 2D TMCs can be classified into two categories:

1. Top-down methods: exfoliation is traditionally used to obtain thin-layer 2D materials from bulk. Mechanical exfoliation (ME) was firstly used to obtain 2D TMCs.[105] Efforts have been made to increase the size of 2D TMCs.[106, 107] For example, Huang et al. used Au-assisted mechanical exfoliation to acquire ~ 1 mm 2D TMCs with some irregular shapes and cracks.[108] It should be noted that larger-scale 2D TMCs are hard to be produced from the ME method, whose scalability and efficiency are low. Therefore, reports on wafer-scale 2D TMCs based on mechanical exfoliation are rare. ~~There is also liquid~~ Liquid exfoliation is also used to obtain thin layered materials, in which solvent molecules, ions, or small organic molecules are used to break down layered structures.[109] Nevertheless, the agitation of liquid exfoliation may damage the completeness of layered structures, usually leading to small size from several hundred nanometers to a few micrometers and generally low monolayer yield. It also result in high density of defects and residual solvent. Thus it currently is not suitable towards industrial applications.[109, 110]

2. Bottom-up methods: atoms can assemble to form a 2D structure, contrasting the thinning of a bulk material down to a monolayer. Efforts in this aspect have led to the largest size for 2D materials to date.[111-113] Typical approaches are chemical vapor deposition (CVD), metal-organic chemical vapor deposition (MOCVD), atomic layer deposition (ALD), pulsed laser deposition (PLD), etc. Among them, CVD prevails due to its simple principle, low-cost setup, easy manipulation and scalability.[114, 115] MOCVD with uniformly and precisely controlled supply of metal-organic precursors is also a potent approach.[116] ALD, which possesses the advantage of atomic precision of controllability and proceeds at low temperatures, is likewise a promising tool to achieve large-area 2D materials with a controllable manner.[117] PLD, featuring the versatility of deposited material, fast growth speed, and excellent ability to retain stoichiometry, has been also reported to achieve wafer-scale 2D TMCs.[118] In total, the scalability, low cost and versatility make bottom-up approaches in the mainstream for the growth of large-area 2D TMCs.

Herein, we introduce the recent progress of the efforts to grow wafer-scale 2D TMCs. Two different definitions about wafer-scale 2D materials growth have been used. One refers to continuous 2D materials covering all the area of a wafer. The other refers to discrete small-area 2D materials scattering over a wafer. In this review we define the wafer-scale 2D TMCs as continuous 2D TMCs whose size is at least in the order of millimeters or larger than the size of devices fabricated from the 2D TMCs, which are distributed over the whole wafer. In this review, extensive recent efforts for growing large-area TMCs is firstly presented. Key factors associated with the quality control of wafer-scale 2D TMCs such as grains, morphology, thickness and phase are addressed. The device applications of wafer-scale 2D TMCs in various areas such as electronics and optoelectronics are then introduced. Finally, we draw conclusions and list future perspects and opportunities on the development of wafer-scale 2D TMCs. Figure 2 exhibis the overview of this review.

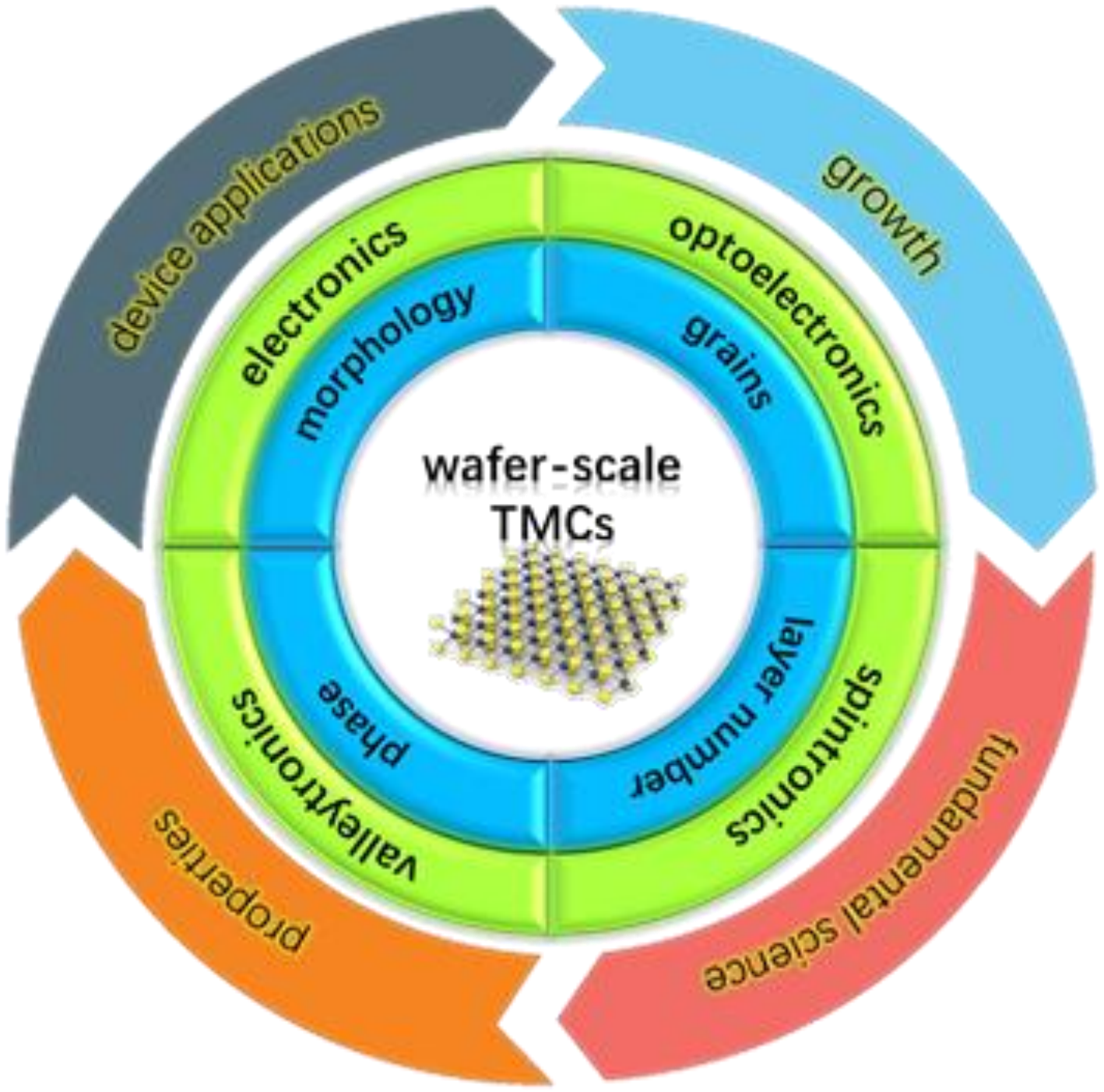

Figure 2. Schematic representation summarizing the content of this review about wafer-scale 2D TMCs.

## 2. Wafer-scale Growth Efforts of Transition Metal Chalcogenides

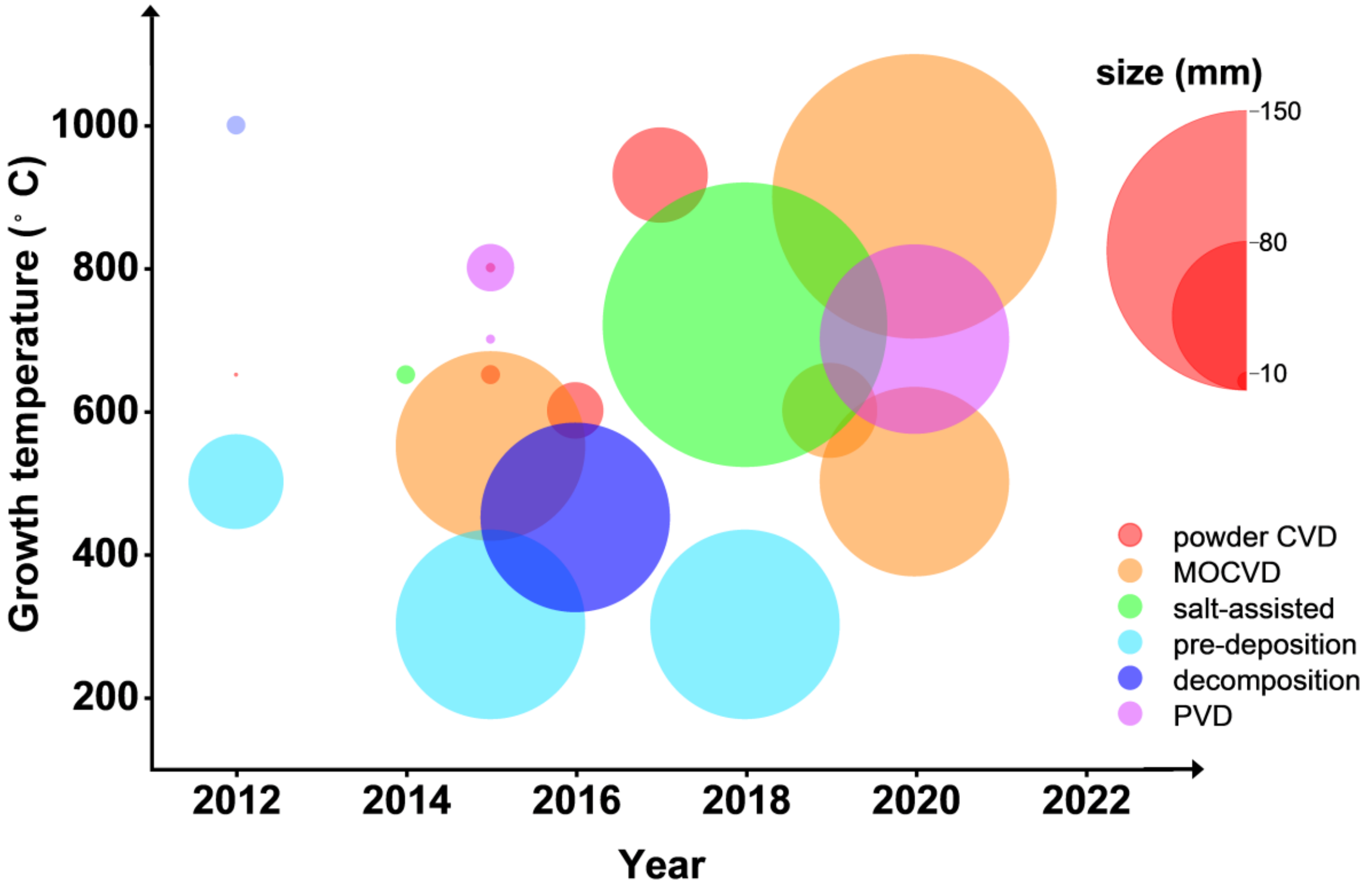

Figure 3. Bubble chart showing the trend of the product size of some representative works in the development of the endeavors for wafer-scale TMCs. The size of the bubbles represents the maximum size for each synthetic approach in millimeter. The y-axis coordinate of the center of circle corresponds to the growth temperature of each work. The right side are the size-scale of the bubbles and color legend.

Transition Metal Chalcogenides (TMCs) possess a descent mobility (up to 200 $cm^2 \cdot V^{-1} \cdot s^{-1}$), high on/off ratios (~$10^8$), low subthreshold slope (70-75 $mV \cdot dec^{-1}$ close to the theoretical limit 60 $mV \cdot dec^{-1}$, monolayer $MoS_2$),[119-121] holding promises for application in the low-standby power integrated circuits. Wafer-scale TMCs material is key to the application in electronics and optoelectronics. Various strategies have been applied to grow wafer-scale TMCs, including conventional CVD with powder precursors, MOCVD, salt-assisted CVD, pre-deposition film, decomposition of precursors, and physical vapor deposition (PVD). Figure 3 intuitively illustrates the trends of the maximum size achieved which is very important for the following device fabrications from some representative works in recent years for the wafer-scale TMCs growth by various approaches and table 1 summarizes the results of these recent representative works for the wafer-scale 2D TMCs and the performance of the as-grown materials (due to the difficulty to control the nucleation, the vast majority of the current results are poly-crystals). Through them, we can gain an genral idea about the progress to produce wafer-scale TMCs in recent years.

**Table 1. Summary of the growth for wafer-scale 2D TMCs materials by various methods**

| materials | methods | maximum size | substrate | growth parameters | performance (mobilities, on/off ratio etc.) | ref |
|---|---|---|---|---|---|---|
| $WS_2$ | powder CVD | 4 cm | sapphire | $1\times10^{-3}$ Torr, 980℃, 30 sccm Ar, 60 mins | mobility: 0.02 $cm^2\cdot V^{-1}\cdot s^{-1}$, photoresponsivity (R): 0.52 mA $W^{-1}$, on/off ratio: of 5.5 $\times 10^3$ | [122] |
| $MoS_2$ | powder CVD with $H_2S$ | 5×2 cm | $SiO_2/Si$ | $10^{-3}$ Torr, 600℃, 200 sccm Ar, 1sscm $H_2S$, 30 mins | 1 $cm^2\cdot V^{-1}\cdot s^{-1}$, on/off ratio: $10^5$ | [123] |
| $MoS_2$ | powder CVD with separate pathways | 5.08cm (2 inch) | sapphire | 1 Torr, 930℃, 100 sccm Ar, 75/3 sccm $Ar/O_2$, 40 mins | 40 $cm^2\cdot V^{-1}\cdot s^{-1}$, on/off ratio: ~$10^6$ | [124] |
| $MoTe_2$ | sulfurization of Mo film | ~cm | $SiO_2/Si$ | $10^{-3}$ Torr, 620℃, 4 sccm Ar, 5 sccm $H_2$, 5 mins | 30 $cm^2\cdot V^{-1}\cdot s^{-1}$, on/off ratio: ~$1\times10^4$ | [125] |
| $SnS_2$ | sulfurization of ALD deposited film | 10.16 cm (4 inch) | sapphire | $Ar/H_2S$ (3.5%) atmosphere , 350℃, 60 mins | 0.02 $cm^2\cdot V^{-1}\cdot s^{-1}$, on/off ratio: ~$1\times10^3$ | [126] |
| $MoS_2$ | sulfurization of Mo film | 10.16 cm (4 inch) | graphene/ $SiO_2/Si$ | $Ar/H_2S$ atmosphere , 300℃, 1.5 hrs | overpotential η 0.43V and Tafel slope 91 $mV\cdot dec^{-1}$ | [127] |
| $MoS_2$, $WS_2$ | MOCVD | 10.16 cm (4 inch) | $SiO_2/Si$ | 7.5 Torr, 550℃, 0.01 sccm $Mo(CO)_6$/ $W(CO)_6$, 0.4 sccm diethyl sulfide, 5 sccm $H_2$, 150 sccm Ar, 26 hrs | 29 $cm^2\cdot V^{-1}\cdot s^{-1}$ ($MoS_2$), 5 $cm^2\cdot V^{-1}\cdot s^{-1}$ ($WS_2$), on/off ratio: ~$10^6$ | [113] |
| $MoS_2$, $WS_2$ | MOCVD | 15.24 cm (6 inch) | $SiO_2/Si$ | 5 Torr, 900℃, 0.024 sccm $Mo(CO)_6$/ $W(CO)_6$, 0.99 sccm diethyl sulfide, 100 sccm $H_2$, 12 mins | 9.4 $cm^2\cdot V^{-1}\cdot s^{-1}$ ($MoS_2$), on/off ratio: ~$10^7$ | [112] |
| $MoS_2$ | organic-alkali metal salt assisted growth | ~mm | $SiO_2/Si$ | $1\times10^{-6}$ Torr, 650℃, 5 sccm Ar, 350℃, 3 mins, 50μM PTAS drop-cast | / | [128] |
| $MoS_2$ | alkali metal salt assisted growth | 15.24 cm (6 inch) | soda-lime glass | low pressure, 720℃, 50 sccm Ar and 6 sccm $O_2$, 8 mins | 6.3~ 11.4 $cm^2\cdot V^{-1}\cdot s^{-1}$ on/off ratio: $10^5$~$10^6$ | [111] |
| $PtSe_2$, $PtTe_2$ | alkali metal salt assisted reaction of the pre-deposited Pt film | 1×1 cm | NaCl | $20\times10^{-3}$ Torr, 400℃, 100 sccm Ar, 50 mins | > $10^6$ S/m electrical conductivities, low on/off ratio | [129] |
| $MoS_2$ | thermal decomposition | ~cm | sapphire or $SiO_2/Si$ | 1 Torr, 4:1 $Ar/H_2$ flow rate, 500℃, 60 mins, 1000℃ annealing in Ar or Ar+S | 4.7 $cm^2\cdot V^{-1}\cdot s^{-1}$, on/off ratio: $1.6\times10^5$ | [130] |
| $MoS_2$ | thermal decomposition | 1×1 cm | polyimide | 1.8 Torr, $N_2$ flow, 280℃, 30 mins; 1.8 Torr, 100 sccm $H_2/N_2$, 450℃, 30 mins | photocurrent decreased 5.6% after $10^5$ cycles of bending | [131] |
| $WSe_2$ | PLD | 1×1 cm | $SiO_2/Si$ | 0.1 Torr, Ar atmosphere, 1 $J\cdot cm^{-2}$, 3 Hz, 5cm distance target /substrate | $5.28\times10^{-3}$ $cm^2\cdot V^{-1}\cdot s^{-1}$, on/off ratio: $1.3\times10^2$ | [132] |
| $MoS_2$ | PLD | 5×5 mm | sapphire, GaN(0001), SiC-6H (0001) | $1\times10^{-6}$ Torr, 700℃, 0.05 $J\cdot cm^{-2}$, 4 Hz, 5cm distance target /substrate | resistivity of 1.6 Ω·cm | [133] |
| $MoTe_2$ | MBE | 5.08cm (2 inch) | $SiO_2/Si$ | 355℃, 100mA $I_{Mo}$, 16 mins | 23.4 $cm^2\cdot V^{-1}\cdot s^{-1}$, on/off ratio: $10^7$ | [134] |

## 2.1 Conventional powder chemical vapor deposition

Due to its easy accessibility, the conventional chemical vapor deposition with the powder precursors positioned in the evaporation boat, is still the most widely used approach. Factors such as temperature, distance between the precursors and the substrate, flow rate etc. need to be optimized to achieve large area growth of TMCs. For example, Lan et al used heating belt located outside the two zone furnace to stabilize the temperature of sulfur powder and grew 4 cm wafer-scale $WS_2$ on sapphire (Figure 4a). [122] The average grain sizes are found to be 300 nm and 2 μm at deposition temperatures of 920℃ and 980℃ respectively, suggesting relative high deposition temperature favors the the synthesis of high-quality crystals. Kim et al. also used solid $MoO_3$ precursors and $H_2S$ gas to grow a continuous $MoS_2$ film on a 5.08cm (2-inch) sapphire substrate (Figure 4b).[123] Moreover, Yu et al. used separate pathways for Mo and S source to grow a continuous $MoS_2$ film on a 5.08cm (2-inch) sapphire substrate (Figure 4c).[124] The independent pathways for the sources ensure a better-controlled feeding of the precursors.Besides, Chen et al. used the substrate of GaN to acquire $MoSe_2$ thin film in a size of 1×1cm. The GaN substrates possess some benefits: the symmetry matches between GaN and $MoSe_2$, and the Ga polarities facilitate the absorption of $Se^-$ anions. The substrate faced down to the below $MoO_3$ source and continuous film was obtained in the zone of the substrate closest to the $MoO_3$ precursor (Figure 4d).[135] Through these reports, the quality of these conventional powder CVD grown samples is comparable to the mechanically exfoliated samples, in terms of PL intensities and full width at half-maximum, mobilities and on/off ratios from field effect transistor (FET) devices. The difficulty of the conventional powder CVD is that it is not easy to avoid overgrowth, discontinuity, and multilayer regions (Figure 4e).[122, 135, 136] Due to the complexity of the growth in the tube furnace, people have to do multiple trials to optimize the condition The repeatability and controllability are not yet excellent.[137, 138]

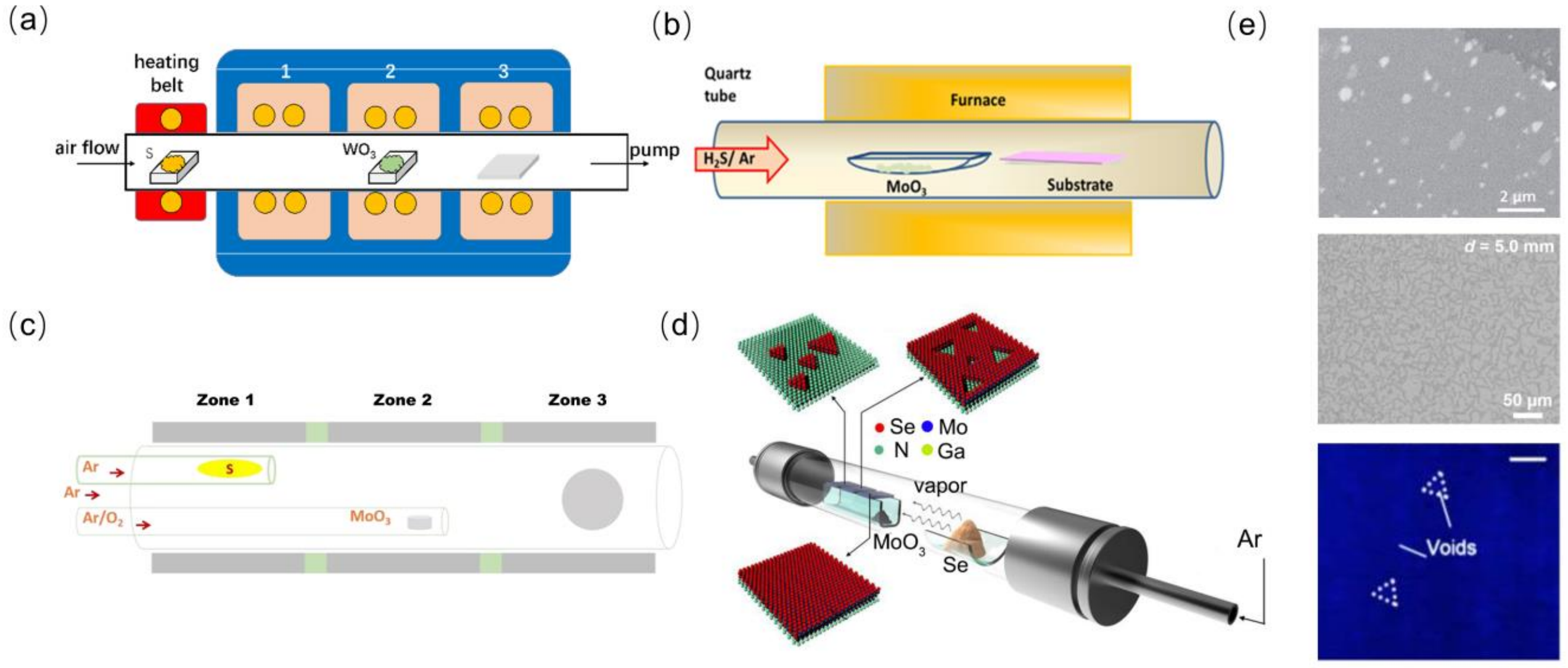

Figure 4. (a-d) The setups and configuration of the precursors and the substrates for the conventional powder CVD to large area TMCs. (e) Overgrowth, voids, and unevenness of the film grown by the conventional powder CVD. (b,e) Reproduced with permission from ref 123. Copyright 2016, IOP Publishing. (c) Reproduced with permission from ref 124. Copyright 2017, American Chemical Society. (d, e) with permission from ref 135. Copyright 2016, American Chemical Society. Reproduced with permission from ref 122. Copyright 2018, Springer Nature.

## 2.2 Sulfurization/selenization/tellurization of the pre-deposited thin film

The second path is to sulfurize/selenize/tellurize the pre-deposited metal or metal oxide films on substrate.[114] Xu et al. prepared a 1-1.5nm thick Mo film on the $Si/SiO_2$ substrate via magnetron sputtering, and tellurized the Mo film at 620℃ with different growth times. 1T’ phase $MoTe_2$ was found to fully form at centimeter scale for the short growth time.[125] Atomic layer deposition (ALD) has the advantage to achieve full coverage and spatial uniformity of the thin film of oxides as precursors.[139, 140] Pyeon et al. used ALD to deposit $SnO_2$ or SnO on a 10.16 cm (4-inch) substrate as mother material, and then sulfurized the tin oxide thin film under Ar/$H_2S$ at a low temperature of 350 ℃ (Figure 5a).[126] The sulfurized film thickness could be controlled by the number of ALD cycles. the product $SnS_2$ had a relatively uniform thickness around 4.9 nm.[126] Kim et al. transferred CVD-grown graphene onto $SiO_2$/Si wafer and deposited a thin film of molybdenum as a seed layer with e-beam evaporation and used plasma-enhanced chemical vapor deposition (PECVD) to sulfurize the Mo-deposited on graphene at 300℃ under $H_2S$ and Ar plasma atmosphere (Figure 5b), obtaining a uniform and continuous $MoS_2$ film with a thickness of 6-7 nm (Figure 5c).[127] Since the precursor is distributed over all the substrate, grains will form and coalesce. The Raman and high resolution transmission electron microscopy (HR-TEM) results revealed that the 5-6 $MoS_2$ layers has a large density of sulfur vacancies and grain boundaries (Figure 5d, e). Note the surface energy of the substrate also has influence on the morphologies of the grown film. For the $SnS_2$ film grown by the sulfurization of SnO, $Al_2O_3$ with high surface energy have a full coverage with the sulfurized layer to minimize the surface energy;[141] whileon the $SiO_2$ surface with low surface energy, the sulfurized layer favors to form discrete islands (Figure 5f, g show atomic force microscopy (AFM) images on different substrates).[126, 142] Meanwhile, a low surface energy due to the lattice type matching and the atomic smoothness of grapheneare conducive to the $MoS_2$ growth.[143, 144] In total, the advantage of this method is better control of the precursor supply and relative low growth temperature routine powder CVD method,  because the precursor is already located on the substrate, not to approach the substrate through evaporation.

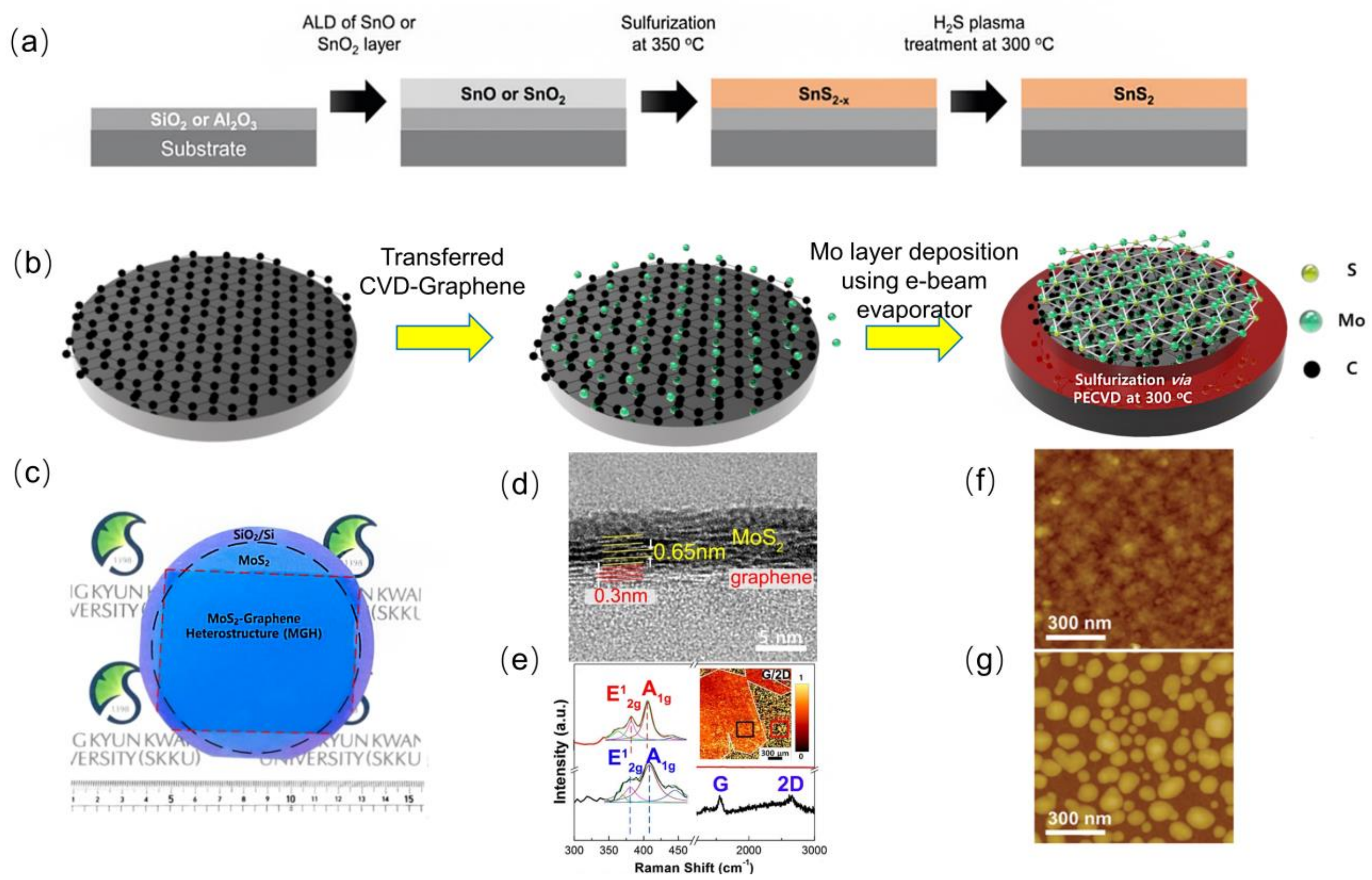

Figure 5. (a) Sulfurization of the pre-deposited SnO or $SnO_2$ film to synthesize $SnS_2$. (b) Sulfurization of the pre-deposited Mo film on graphene that is transferred onto $SiO_2$/Si to prepare $MoS_2$ film. (c) Optical microscope image of the large area $MoS_2$ grown on graphene/$SiO_2$/Si. (d) HR-TEM image of the $MoS_2$ grown on graphene/$SiO_2$/Si. (e) Raman spectra of the $MoS_2$. (f, g) AFM image of $SnS_2$ grown by sulfurization of pre-deposited SnO film on amorphous $Al_2O_3$ (f) and on $SiO_2$ (g). (a, f, g) Reproduced with permission from ref 126. Copyright 2018, Royal Society of Chemistry. (b, c, d, e) Reproduced with permission from ref 127. Copyright 2018, Elsevier.

## 2.3 Gaseous phase precursor method

The third path is to use gaseous phase precursor. Metal-organic chemical vapor deposition (MOCVD) is a typical approach to use all gaseous phase precursors (for example, $W(CO)_6$ as transition metal precursor and $H_2S$ as chalcoge precursor). The supply can be readily switched on/off by the gas flow and the amount of supply of precursors can be adjusted by the flow rate.[145] MOCVD usually holds the ability to process a whole wafer substrate.

For example, Kang et al. used molybdenum hexacarbonyl ($Mo(CO)_6$), tungsten hexacarbonyl $W(CO)_6$ as the molybdenum/tungsten precursor and ethyl sulfide $(C_2H_5)_2S$ as the sulfur precursor, and synthesized $MoS_2$ and $WS_2$ on a 10.16 cm (4-inch) wafer (Figure 6a, b, c).[113] Devices based on these materials show a median field effect mobility of 5 $cm^2{\cdot}V^{-1}{\cdot}s^{-1}$ and high on/off ratio of $10^6$. Very recently, Seoul et al. used a pulsed MOCVD, which included periodic interruption in the precursor supply inside a shower-head

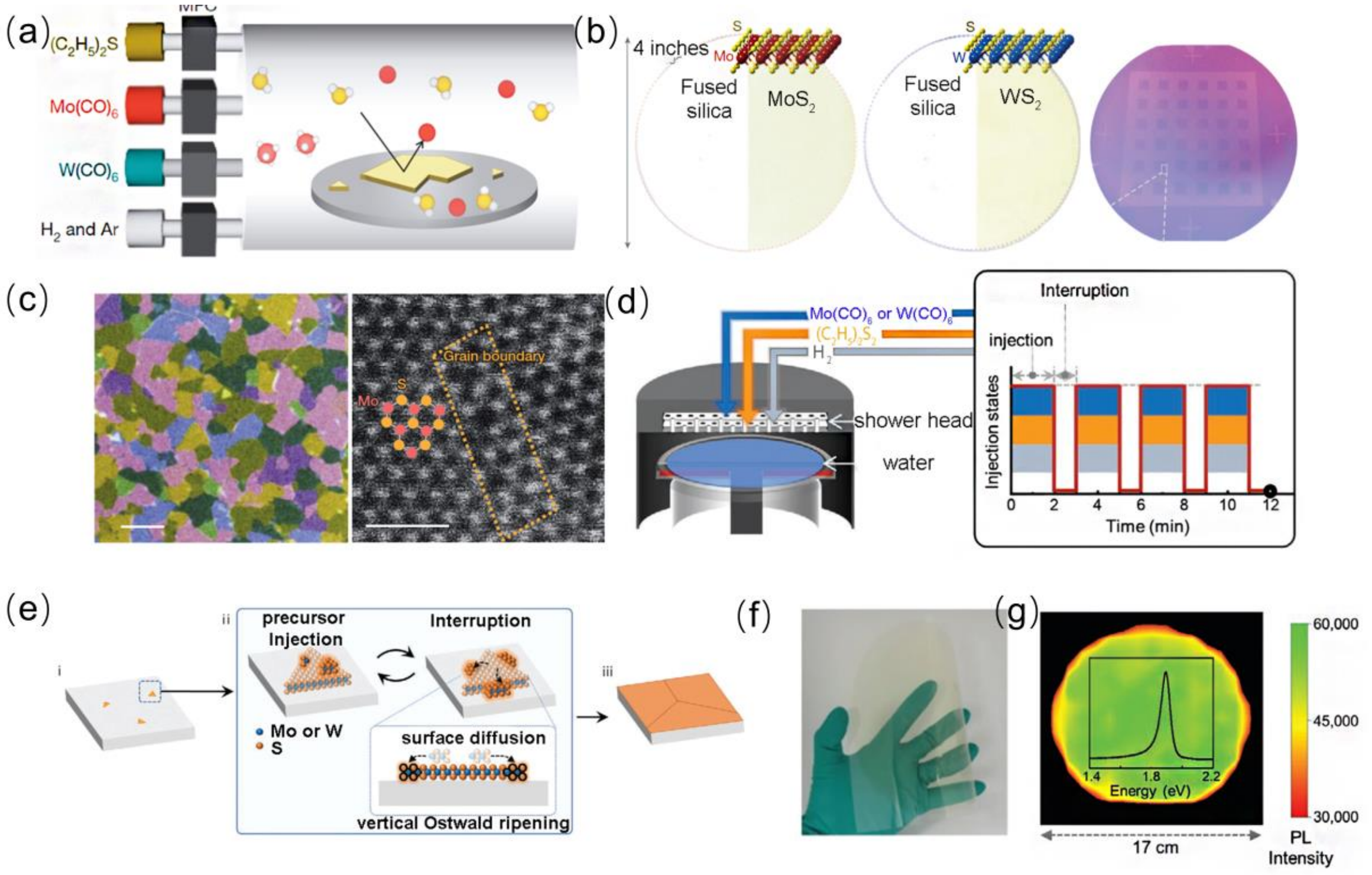

Figure 6. (a) Schematics of the MOCVD growth of $MoS_2$ and $WS_2$. (b) $MoS_2$, $WS_2$ grown on fused silica and Si. (c) False color darkfield TEM images showing the stitching of the $MoS_2$ and the annular darkfield scanning TEM showing the grain boundaries of the $MoS_2$ film. (d) The pulsed MOCVD with periodic interruption in the precursor supply. (e) Vertical Ostwald ripening in the pulsed MOCVD. (f) 15.24cm (6-inch) $MoS_2$ film transferred onto flexible polyethylele terephthalate (PET). (g) PL intensity mapping of the $MoS_2$ film, showing the uniformity. (a, b, c) Reproduced with permission from ref 113. Copyright 2015, Springer Nature. (d, e, f) Reproduced with permission from ref 112. Copyright 2020, Wiley-VCH.

type cold-wall reactor system (Figure 6d), purposefully to suppress the secondary nucleation, vertical growth by the vertical Ostwald ripening and relaxation on the edge of $MoS_2$ (Figure 6e). They achieved the growth of a monolayer $MoS_2$ film with high uniformity on a 15.24cm (6-inch) $SiO_2$/Si wafer (Figure 6f, g).[112] The wafer-scale homogeneous TMCs film allows back-end-of-line integration into Si CMOS platform, displaying the advantage of gaseous phase precursors.[146, 147]

Moreover, growth in CVD tube furnace with gaseous phase precursors is also beneficial for achieving large area material. Tang et al. reported to substitute the widely used solid precursors with gaseous precursor ($H_2S$ and Ar-bubbled metal precursor feed) in a vertical CVD system and produced spatially highly uniform $10 \times 10$ mm$^2$ continuous $WS_2$ over both sapphire and $SiO_2$/Si substrates which showed good stoichiometry and quality.[148] The controllable and steady concentration offered by the gaseous phase precursors helped to overcome the issue of the gradient precursor concentration, resulting in a steady mass flux transportation into the growth chamber and also a uniform growth profile. The redistribution of the temperature and gas field due to the change in the orientation of the furnace which helps to eliminate posisition dependence contributed to more uniform growth as well.

In summary, using gaseous phase precursor has the advantage of more precise controllability and highly uniform distribution of the feeding of the precursor. The limit is that the grain size of growth results is somewhat small (tens of nm), and the rich grain boundaries reduce the electrical properties.

## 2.4 Alkali metal salt-assisted growth

In 2018, the well-known Nature paper about that the alkali metal salt lowers the melting points of the metal or metal oxide precursors and comprehensively benefits the TMC growth widely promotes the salt-assisted growth. A library of 2D metal chalcogenides has been synthesized (Figure 7a).[149] Since then, the alkali-metal salt has been broadly applied to synthesize 2D TMCs. It is found that alkali-metal salt and transition metal oxide precursor can form eutectic intermediate, lowering the energy barrier for the generation of TMCs.[149-152] For example, for the $MoO_{3-x}$ reacting with sulfur, after adding alkali-metal salt, $MoO_{3-x}$ undergoes the following reactions:[153]

$$M_aX + MoO_{3-x} \rightarrow M_2MoO_4 + M_2Mo_2O_7 + \text{byproduct}\uparrow \qquad \text{eq.1}$$

$$2M_2MoO_4 + 7S \rightarrow 2MoS_2 + 2M_2O + 3SO_2 \qquad \text{eq.2}$$

$$M_2MoO_4 + 5S \rightarrow MoS_2 + M_2S + 2SO_2 \qquad \text{eq.3}$$

where M refers to alkali metal elements (Na, K, and Li), while X refers to anions. The eutectic intermediates possess low melting point. Both reactions of eq.2, 3 have extremely large negative Gibbs free energy ($\Delta G$), thus being favored thermo-dynamically.[153] This helps to form TMCs.

Yang et al. made use of the Na in the soda-lime glass substrate. A Mo foil was place above the soda-lime glass substrate as Mo source (Figure 7b), instead of $MoO_3$. With the aid of $O_2$ to convert Mo to chemically active $MoO_{3-x}$, a continuous monolayer $MoS_2$ film was grown on a diagonal 15.24cm (6-inch) rectangular soda-lime glass (Figure 7c). Na widely distributed in the confined space between Mo foil and glass during the CVD growth process and served as an intermediate catalyst. Furthermore, density functional theory (DFT) calculation revealed that with the incorporation of Na, the highest energy barrier (from $Mo_2S_2$ to $Mo_2S_4$) reduces from 0.53 to 0.29 eV (Figure 7d, e), favoring the reaction.[111]

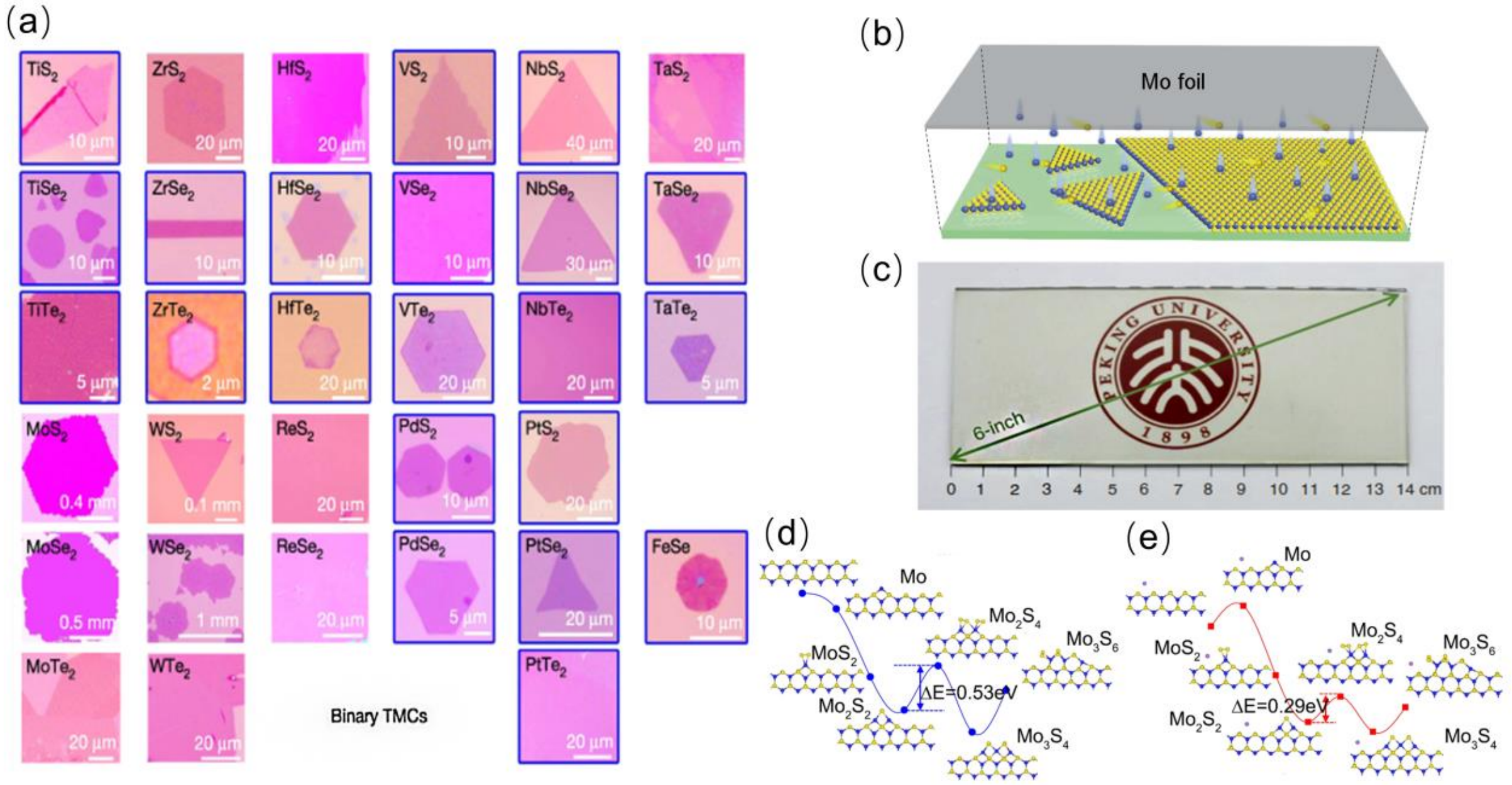

Figure 7. (a) A library of TMCs grown by assistance of salt. (b) The schematic showing the face-to-face configuration of the Mo foil and the soda-lime glass. (c) 15.24 cm (6-inch) $MoS_2$ film synthesized on soda-lime glass for ~ 8 min. (d) the DFT-calculated energy diagrams for $MoS_2$ growth along the S-terminated edges, without (d) and with (e) Na adsorption. The energy barrier changes from 0.53eV to 0.29eV. (a) Reproduced with permission from ref 149. Copyright 2018, Springer Nature. (b, c, d, e) Reprrroduced with permission from ref 111. Copyright 2018, Springer Nature.

Moreover, lately, Han et al. deposited Pt thin film on the judiciously selected alkali metal salt substrates (NaCl, KCl, KBr) to grow 2D $PtSe_2$ and $PtTe_2$. The size of the continuous film reached 1cm×1cm (Figure 8a). The temperature for growth was 400℃ due to the factor the alkali metal atom effectively lowers the melting point of the transition metal precursor. The water-soluble feature of the salt substrate highlights the feasibility to conveniently transfer the film with the assistance of water (Figure 8b).[129]

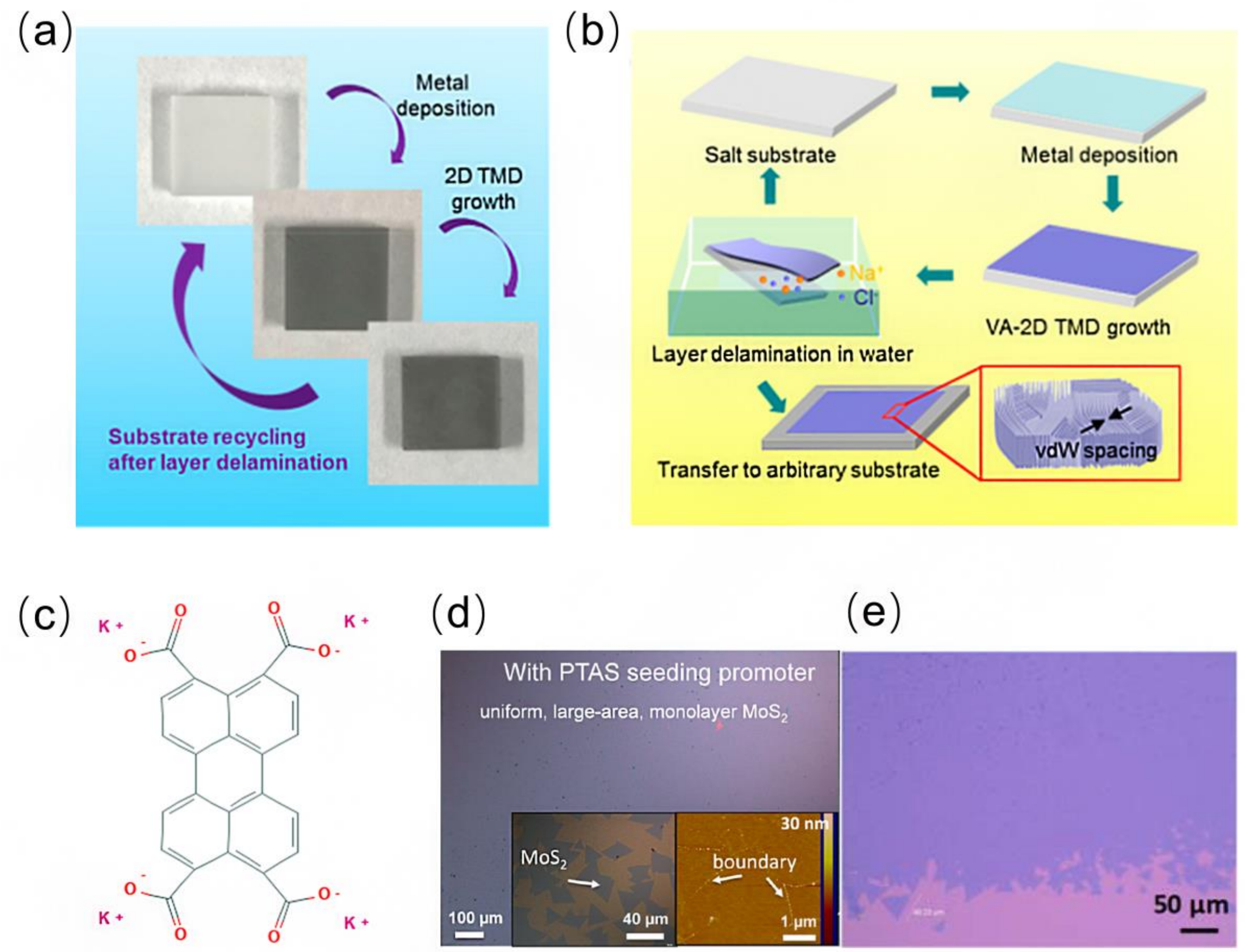

Figure 8. (a) Sulfurization of the Pt film deposited on the salt substrate (NaCl, KCl, KBr) to grow $PtSe_2$ and $PtTe_2$. (b) Schematic showing the salt substrate-assisted growth and the transfer of the layer in water afterwards, exploiting the water soluble property of the substrate. (c) The molecular structure of the organic-alkali salt PTAS. (d-e) The $MoS_2$ film grown by the PTAS-assisted CVD. (a, b) Reproduced with permission from ref 129. Copyright 2020, American Chemical Society. (d) Reproduced with permission from ref 128. Copyright 2014, American Chemical Society. (e) Reproduced with permission from ref 154. Copyright 2018, IOP Publishing.

Furthermore, perylene-3,4,9,10-tetracarboxylic acid tetrapotassium salt (PTAS), an organic-alkali salt (Figure 8c), was drop-cast or spun-cast in solution onto the $SiO_2$/Si substrate by Ling et al. (Figure 8d) and Cai et al. (Figure 8e), and millimeter-scale continuous film was attained.[128, 154] The organic-alkali metal salt can function primarily as the catalyst, and secondly as the seed promoters, leading to an enhanced vapor pressure of the transition metal precursor, which benefits the large area TMCs growth.[155, 156]

The advantage of the salt assisted growth is the growth can be achieved at relatively low temperature because melting point of the metal precursor can be lowered. But some alkali metal salt maybe remain as residues on the grown samples.[116, 157]

## 2.5 Thermolysis of the pre-coated transition metal-chalcogen salt

Reversed to the common strategy of the reaction of the transition metal precursor and chalcogen precursor, an interesting strategy to decompose pre-coated transition metal-chalcogen-ammonia compound was reported. Liu et al. prepared solution of $(NH_4)_2MoS_4$ in dimethylformamide (DMF) with 1.25 wt % and dip-coated the substrate such as sapphire or $SiO_2$/Si to form a thin $(NH_4)_2MoS_4$ film. Subsequently, they used a two-step process to obtain $MoS_2$ film: first, annealed the substrate in Ar/$H_2$ flow at 500°C, 1 Torr for 1 hour. Then Liu annealed the substrate in the mixture of Ar and sulfur at 1000°C, 500 Torr for another 30 mins (Figure 9a). The following reaction occurred:

$$(NH_4)_2MoS_4 + H_2 \rightarrow 2NH_3 + 2H_2S + MoS_2 \qquad \text{eq.4}$$

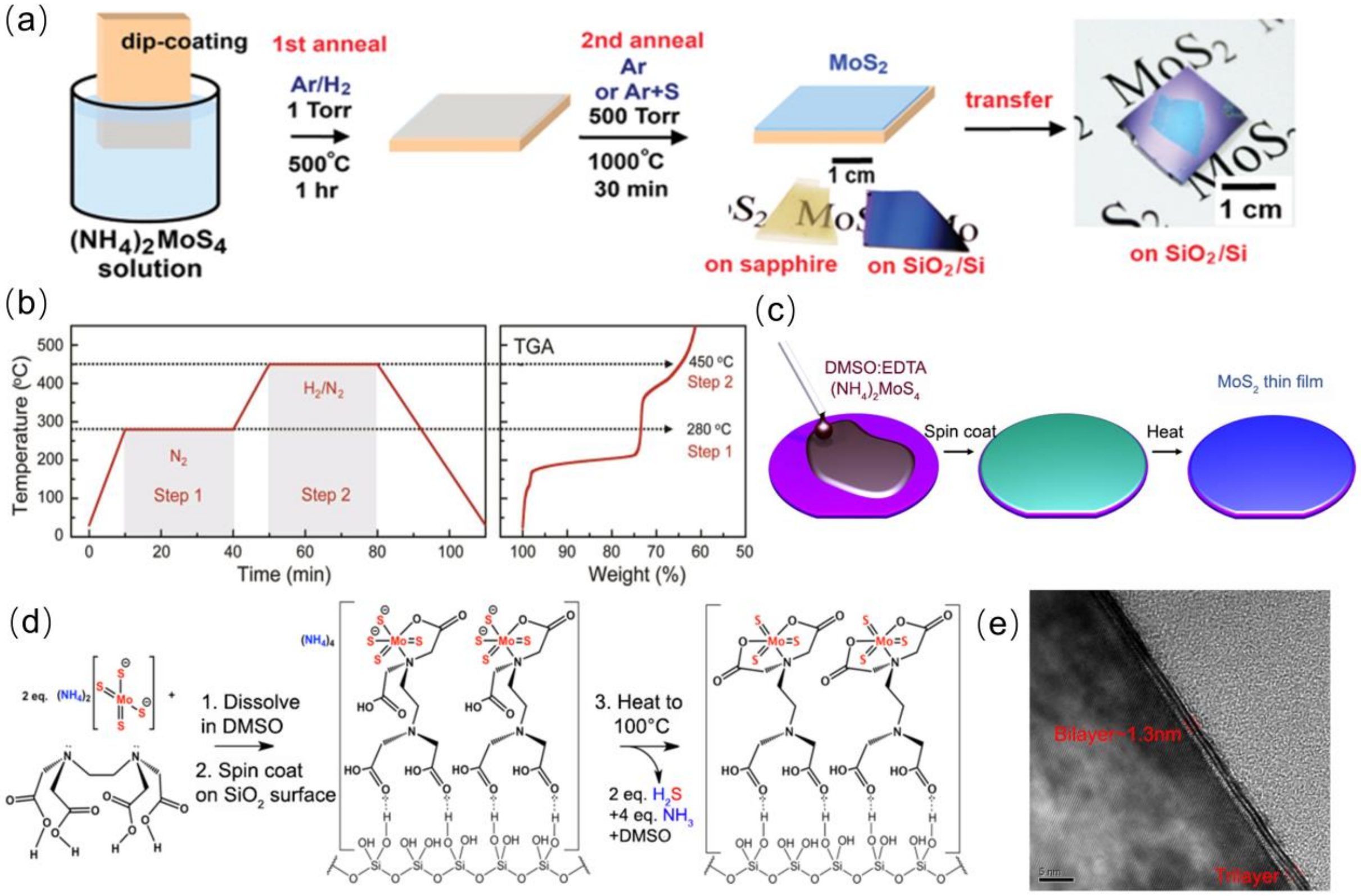

Figure 9. (a) Two step growth of $MoS_2$ by annealing the substrate dip-coated by $(NH_4)_2MoS_4$ solution. ~1cm scale $MoS_2$ was acquired. (b) The temperature profile for the growth of $MoS_2$ on polyimide film. (c) The schematic showing the preparation of the wafer-scale $MoS_2$ film by spin-coating $(NH_4)_2MoS_4$ in the DMSO/EDTA chelant solution and then annealing. (d) The chemistry to coordinate the precursor onto the hydroxyl-functionalized $SiO_2$/Si by the chelant. (e) TEM image of the centimeter-scale $WS_2$ prepared by decomposition of $(NH_4)_2WS_4$. (a) Reproduced with permission from ref 130. Copyright 2012, American Chemical Society. (b) Reproduced with permission from ref 131. Copyright 2016, Wiley-VCH. (c, d) Reproduced from ref 158 under the terms of the Creative Commons Attribution 4.0 International License (http://creativecommons.org/licenses/by/4.0/). Copyright 2017, Springer Nature. (e) Reproduced with permission from ref 159. Copyright 2020, Springer Nature.

A continuous film of size around 1cm×1cm was obtained. The FET device test showed the field-effect electron mobility was enhanced from the order of $10^{-2}$ to 4.7 $cm^2 \cdot V^{-1} \cdot s^{-1}$ after annealing in sulfur[130] Similarly, Lim et al spun cast $(NH_4)_2MoS_4$ in ethylene glycol onto 1cm × 1cm polyimide film or 10.16 cm (4-inch) $SiO_2$/Si and used heat to obtain wafer-scale, homogeneous and stoichiometric few-layer $MoS_2$ (Figure 9b, c).[131] Ionescu et al used chelant ethylenediaminetetraacetic acid (EDTA) in Dimethyl-sulfoxide (DMSO) to coordinate $MoS_4^{2-}$ anions to hydroxyl-functionalized surface of $SiO_2$/Si substrate and heated it to obtain $MoS_2$ film (Figure 9d).[158] Likewise, Abbas et al. spun-coat $(NH_4)_2WS_4$ solution in n-methylpyrrolidone, n-butylamine, 2-aminoethanol mixture and thermally decomposed it to obtain continuous and uniform 2~3 layer $WS_2$ film over centimeter scale (Figure 9e), with the field effect mobilities varying from $10^{-4}$ to $10^{-2}$ $cm^2 \cdot V^{-1} \cdot s^{-1}$. [159] The homogeneity of the films by this approach is good but the quality is still not high enough, likely due to the transformation process involves many steps which may be easily affected by the prescence of oxygen.[130]

## 2.6 Physical Vapor Deposition

Physical vapor deposition (PVD) remains the most straightforward approach to deposit a thin film.[160, 161] However, because of the defects etc. during the deposition process, it easily leads to the growth mode of Volmer-Weber (VW) mode (3D island) or the Stranski-Krastanov (SK) mode (layer-island), not the Frank-Van der Merwe (FM) layer by layer manner.[51, 162-164] Some techniques have been reported to be applied to the growth of the continuous 2D film. Pulsed Laser Deposition (PLD) is a technique, which deposits the congruent moving energetic particles from a target to a substrate, generated from a plume by the direct injection of pulsed laser onto a target, not involving any chemical reaction. For the advantageous side of PLD growth, the temperature is relatively low (~500℃), compared to the common thermal deposition and electron beam deposition; the fabrication rate is relatively high; and the homogeneity is good. The thickness can be tuned by regulating the number and power of the laser pulses (Figure 10a). Especially, the stoichiometry of the deposit materials could be retained well because the target species are transferred congruently through laser ablation, in the ultrahigh vacuum.[118, 165, 166] Seo et al. reported growth of continuous $WSe_2$ film through PLD, in the scale of 1cm×1cm. They varied the number of pulses, and obtained various thickness from monolayer to trilayers (Figure 10b) with a high uniformity and homogeneity, demonstrated by the Raman mapping of the $E^1_{2g}$ and $A_{1g}$ modes and the root-mean-square (RMS) roughness of 0.26~0.29nm. The mobility and on/off current ratio extracted from the transfer characteristics were $5.28 \times 10^{-3}$ $cm^2 \cdot V^{-1} \cdot s^{-1}$ and $1.3 \times 10^2$ [132] Serrao et al. also used PLD to grow 1-15 layer $MoS_2$ on $Al_2O_3$ (0001), GaN (0001), and 6H-SiC (0001) in size of 5×5 mm, with the thickness regulated by the number of laser pulses (Figure 10c).[133] Godel also used PLD to grow monolayer $WS_2$ on STO and multilayer $WS_2$ on Ni in a size of 1cm×1cm, with a stoichiometric W/S ratio of 0.5 reserved.[167] If the laser spot is fixed, the film will be thicker on the spot the plume and thinner far away, also with the growth size limited. Very recently, Juvaid et al. used a laser beam to raster scan on the target and achieved coverage of $MoS_2$ and $WS_2$ on 10.16 cm (4-inch) fused silica substrates (Figure 10d).[168] From all these experimental results, the uniformity of the PLD grown film is good, but the crystalinity still calls to be improved.In addition, Molecular Beam Epitaxy (MBE) has also been testified to scalably grow wafer-scale TMC films. MBE usually takes place in ultra-high vacuum ($10^{-8}$-$10^{-12}$ Torr), with a very slow injection and deposition rate of the material species, thus to epitaxially grow highly uniform and crystalline films. The advantage of this approach is that ultra-high vacuum environment allows high purity and quality of the films. In-situ test technique such as reflection high-energy electron diffraction (RHEED) could be incorporated to monitor and study the growth of the crystal layers. The disadvantage is that the growth rate is very slow (< 3000 nm per hour) and the cost of the setup is high.[169, 170] He et al. reported to use MBE to scalably grow 2H-$MoTe_2$ on 5.08cm (2-inch) $SiO_2$/Si wafer, gaining a 100% coverage (Figure 10e). The mobilities determined from 100 devices fell in a narrow range of 19.2–27.0 $cm^2 \cdot V^{-1} \cdot s^{-1}$, comparable with exfoliated flakes, suggesting the relative high-quality uniformity.[134] Liu et al also reported to grow wafer-scale 8nm 2D ferromagnetic $Fe_3GeTe_2$ films on sapphire by MBE, with confirmation of a van der Waals gap of 0.82 nm by HR-TEM measurements and indication of high-crystalline quality by RHEED pattern.[171] This shows the versatility of the TMC material that MBE is able to grow.

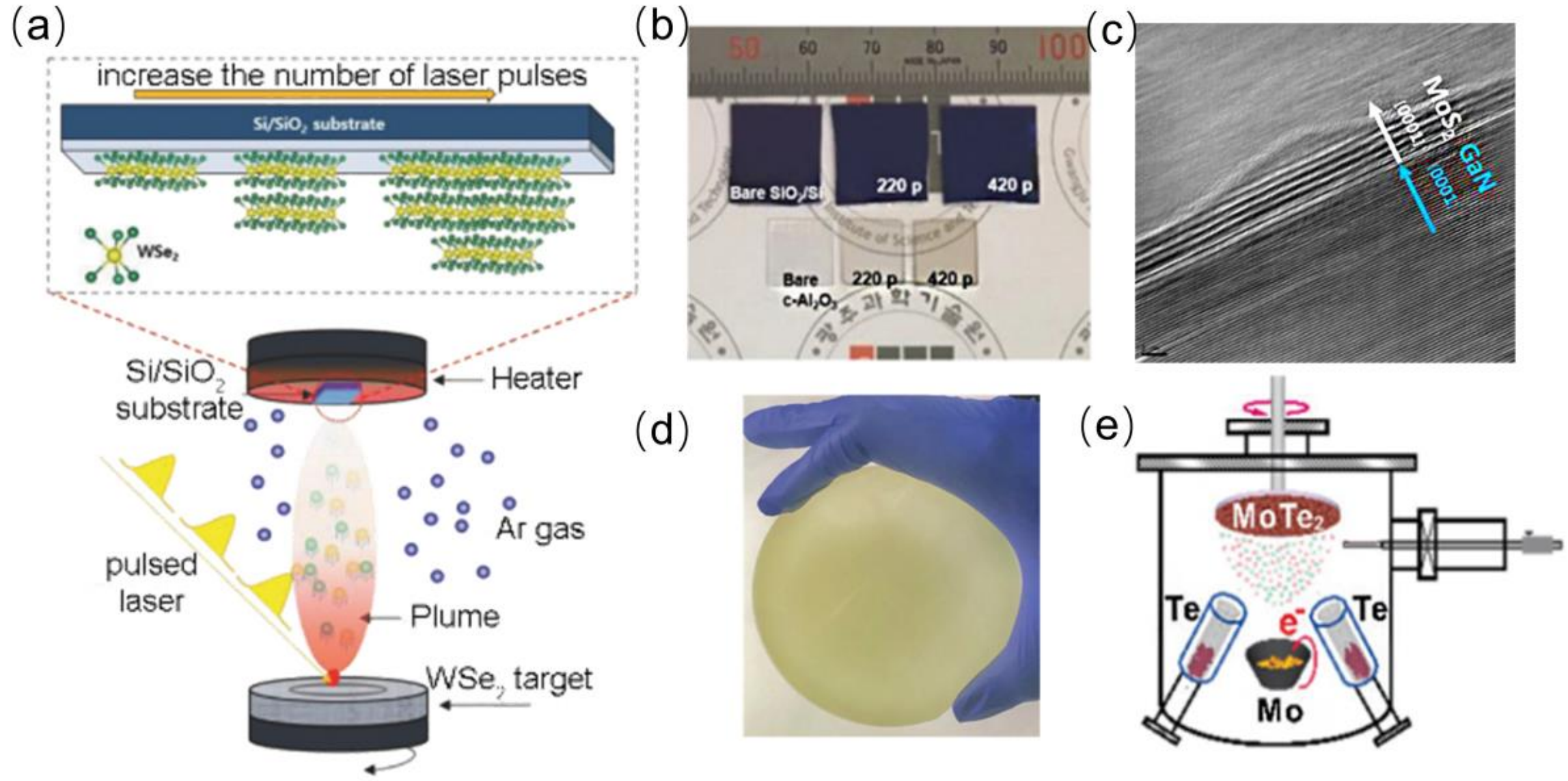

Figure 10. (a) Schematic diagram of deposition of $WSe_2$ by PLD. (b) $WSe_2$ thin films deposited on bare $SiO_2$/Si and c-sapphire with different pulse number. (c) Cross-sectional TEM image of $MoS_2$ grown on GaN(0001). (d) $MoS_2$ grown on a 10.16 cm (4-inch) fused silica by raster-scanning PLD. (e) Schematic diagram of the MBE setup to grow $MoTe_2$. (a, b) Reproduced with permission from ref 132. Copyright 2018, Wiley-VCH. (c) Reproduced with permission from ref 133. Copyright 2015, AIP Publishing. (d) Reproduced with permission from ref 168. Copyright 2020, Elsevier. (e) Reproduced with permission from ref 134. Copyright 2019, Wiley-VCH.

# 3. Factors to control

## 3.1 Grains and morphology

Large area continuous TMCs film is usually stitched by the grains. Annular darkfield scanning TEM image with false color in Figure 6c shows the stitching of the domains. The coalescence of the domains influences the morphologies in a great deal. If the grain size is too small and coalescence is not fully realized, the large area TMCs film will not be continuous, with quite a few voids. Furthermore, when the grain size increases from 20-600 nm, while Raman spectra nearly showed no alteration in peak position and intensity, PL shift is found to shift, resulting from the band-gap modulation (Figure 11a,b).[172] Moreover, grain boundaries, enhanced quantitatively when the grain size reduces, will impair the carrier mobility because of the scattering of the carriers (Figure 11c).[173-175] Therefore, control over the domain size and grain boundaries are important in the large-area TMCs growth, in order to optimize the performance of the TMCs material.[176, 177]

It is well acknowledged that the grain size is closely related to the nucleation density.[41, 178, 179] Thus suppressing the nucleation density, which links with the precursor supply in the mass transport, is essential in order to increase the size of the growth. The growth pressure which influences diffusivity of precursors and the temperature which influences the vapor pressure of precursors can also somehow affect the precursor supply in the mass transport.[180] On the other hand, if precursor supply is insufficient, it will also decrease the size of the product (Figure 11d).[136] Therefore, it is a balance between the suppression of the nucleation density and a sufficient supply of the precursor. In the conventional powder CVD growth, the solid precursor is located in the upstream or below the substrates. It calls for the researchers to adjust the amount the precursor, distance between the precursor and substrates, temperature, flow rates, etc. and optimize them. Researchers devoted much effort in this aspect.[104, 137, 138, 181-186] However, it is purely empirical to make the above balance and the controllability is limited.

Recently, Lim et al. used a NiO foam barrier to suppress the nucleation density by forming Ni-Mo complex, attaining a centimeter-scale continuous monolayer film with the grain sizes from 6-55 um (Figure 11e).[187] Since the vapor pressure of the transition metal oxide (such as $WO_3$) is low,[188, 189] researchers have found the seed promoters such as perylene-3,4,9,10-tetracarboxylic acid tetrapotassium salt (PTAS) facilitates to decrease the surface energy for lateral growth and also stabilize the nuclei for the growth.[128, 190] However, the high density of the seed promoter also resulted in high nucleation density and small domains. Research about optimization or manipulation of the seed promoter density for the wafer-scale TMCs production is still lacking.

Moreover, alkali-metal salt helps the metal precursor to form eutectic intermediate which has a lower melting point. Thus it can increase vapor pressure of the metal precursor and further the grain size of the growth result (Figure 11f). Researchers used soda-lime glass, which possess sodium intrinsically, for large area material growth. For better control of the nucleation density, Mo foil was used. Mo's melting point is 2623℃ and $MoO_3$'s melting point is 795℃. Thus Mo has a lower vapor pressure. Yang et al used small amount of $O_2$ to oxidize the Mo foil for releasing chemically active $MoO_{3-x}$, more controllable than direct evaporation of much more evaporative $MoO_3$ powder. The suppressed nucleation densities and all-over distribution of sodium in the soda-lime glass, leading to a relatively uniform growth of $MoS_2$ and larger grain size (Figure 11g, h).[111] However, research about how the sodium distribution in the soda-lime glass correlates with the growth is still lacking. Likewise, Kim et al studied about using alkali metal halides in metalorganic chemical vapor deposition of $MoS_2$ monolayers and found that alkali metal halides play a dual role in facilitating adsorption of Mo in the pre-exposure stage and suppressing nucleation in the growth stage.[116]

Furthermore, researchers also used other strategies. Higher growth temperature generally increase the mobility of the precursor above the substrate and generates a larger grain size due to the reduced nucleation density.[191-193] Chen et al. found that the "molten glass substrate" creates a quasi-atomic smooth and homogeneous liquid surface, with a lowered migration barrier to decrease nucleation density, allowing an enhanced grain size (Figure 11i).[194] Moreover, introducing a small amount of oxygen into the CVD chamber was proposed to suppress the nucleation density by etching away the unstable nuclei.[116, 195, 196] The escence of all these strategies is to lower the nucleation density.

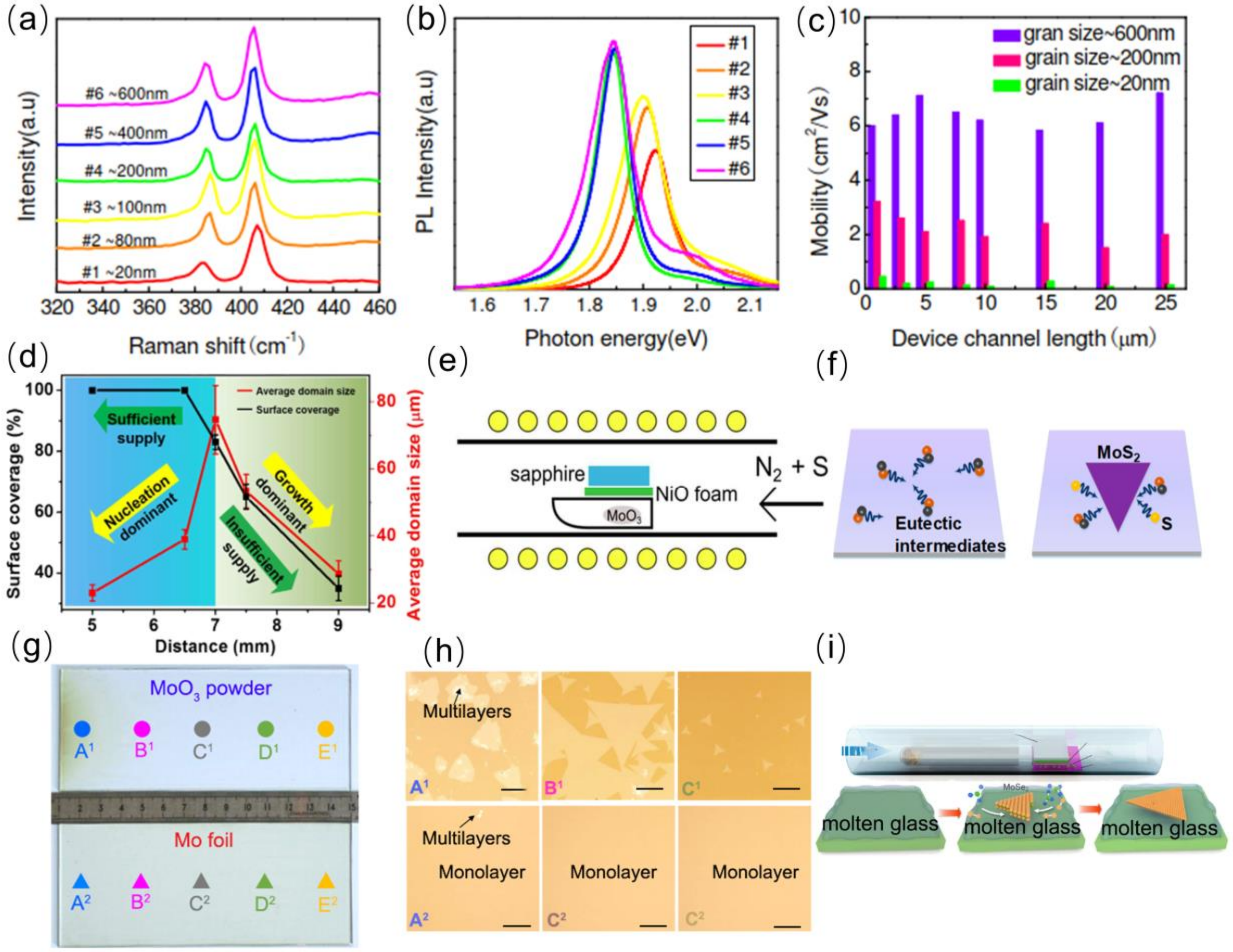

Figure 11. (a) (b) The variation of Raman (a), PL (b) spectra when the grain size increases from 20-600 nm. (c) The variation of mobilities of $MoS_2$ for the grain size of 20, 200, 600 nm. (d)The surface coverage and the average domain size versus distance between the source and the substrate. (e) The growth of $MoS_2$ with suppression of nucleation densities by a NiO foam. (f) The schematic showing the low melting point eutectic intermediates favoring the growth of $MoS_2$. (e)(f) The growth result by the $MoO_3$ or Mo foil as Mo source and corresponding optical microscope images. The approach of Mo foil oxidized with oxygen shows an evener nucleation densities. (g) The schematic showing the idea of using molten glass as the growth substrate. (a, b, e, f, i) Reproduced with permission from ref 172, 187, 194. Copyright 2014, 2018, 2017, American Chemical Society. (c) Reproduced from ref 136, under the terms of the Creative Commons Attribution 4.0 International License (http://creativecommons.org/licenses/by/4.0/). Copyright 2015, Springer Nature. (g, h) Reproduced with permission from ref 111. Copyright 2018, Springer Nature.

Substrates may also play an important role in the growth of the TMCs. Recently, Dong et al. carried out extensive DFT calculations. They theoretically proved that the epitaxial growth of a 2D single-crystal can be realized only if the symmetry group of the substrate is a subgroup of that of the 2D material. The propagation of the edge of a 2D material tends to align with a high symmetry direction of the substrate. This provides guidance for the seamless stitching of the domains and enhancement of the grain size.[197] Aljarb et al. also found experimentally that the trigonal $MoS_2$ tends to align with the hexagonal sapphire in 0° or 60° under high sulfurization in the seeding stage.[198] The studies about the control about the grain propagation in the wafer scale 2D TMC growth are still very rare. If the propagation can be in a well-aligned manner, the grains in the film will be more ordered rather than chaotic, and also the grain size can be enhanced.

It still awaits more investigation to lower the nucleation densities, increase the grain size and control the stitching of the grains in the wafer-scale TMCs growth.

## 3.2 Layer number

The properties of the TMCs material are directly related with the layer number.[199, 200] For the majority of the TMCs materials, the band gap decreases with the increasing of the number of layers.[201, 202] For the application of the wafer-scale TMCs material, the layer number homogeneity is highly desired because in the potential applications of electronic and optoelectronic devices, the homogeneity and uniformity of the device performance is important.[146, 152, 203-205] It roots in the homogeneity and uniformity of the material. One factor that influences the uniformity of the layer number largely is the uniformity of the precursor feeding.[206-208] In the conventional powder CVD, it is a “point-to face” metal-source supply with the metal oxides precursors placed in the upstream, or below the substrate.[209] It is extremely hard to control the uniformity because the inhomogeneous release rates of the metal precursor in different direction and low repeatability due to the time-to-time difference. Then a “face-to-face” configuration was proposed.[111, 147, 210] In the growth of the diagonal 15.24 cm (6-inch) uniform monolayer on soda-lime glass, a Mo foil was placed above the substrate as a bridge to ensure a homogeneous supply of the Mo precursor (Figure 12a, b).[111] Placing a substrate deposited with the transition metal precursor below the growth substrate as a source was proposed as well, with the same idea for a “face-to-face” configuration.[211, 212]

Pre-deposition of the precursor is another approach to offer a more uniform precursor supply.[114, 145, 152, 213] In addition to the conventional thermal evaporation, electron beam deposition or magnetron sputtering,[125, 214-217] atomic layer deposition (ALD) affords atomic film thickness precision control of the deposited film, benefiting the following reaction with chalcogens.[117, 218-220] Tan et al. used ALD to deposit $MoCl_5$ film and sulfurize it with $H_2S$ to generate 5.08 cm (2-inch) continuous film.[140] ALD provides a route to tune the layer number of the product through deposition cycles (Figure 12c).

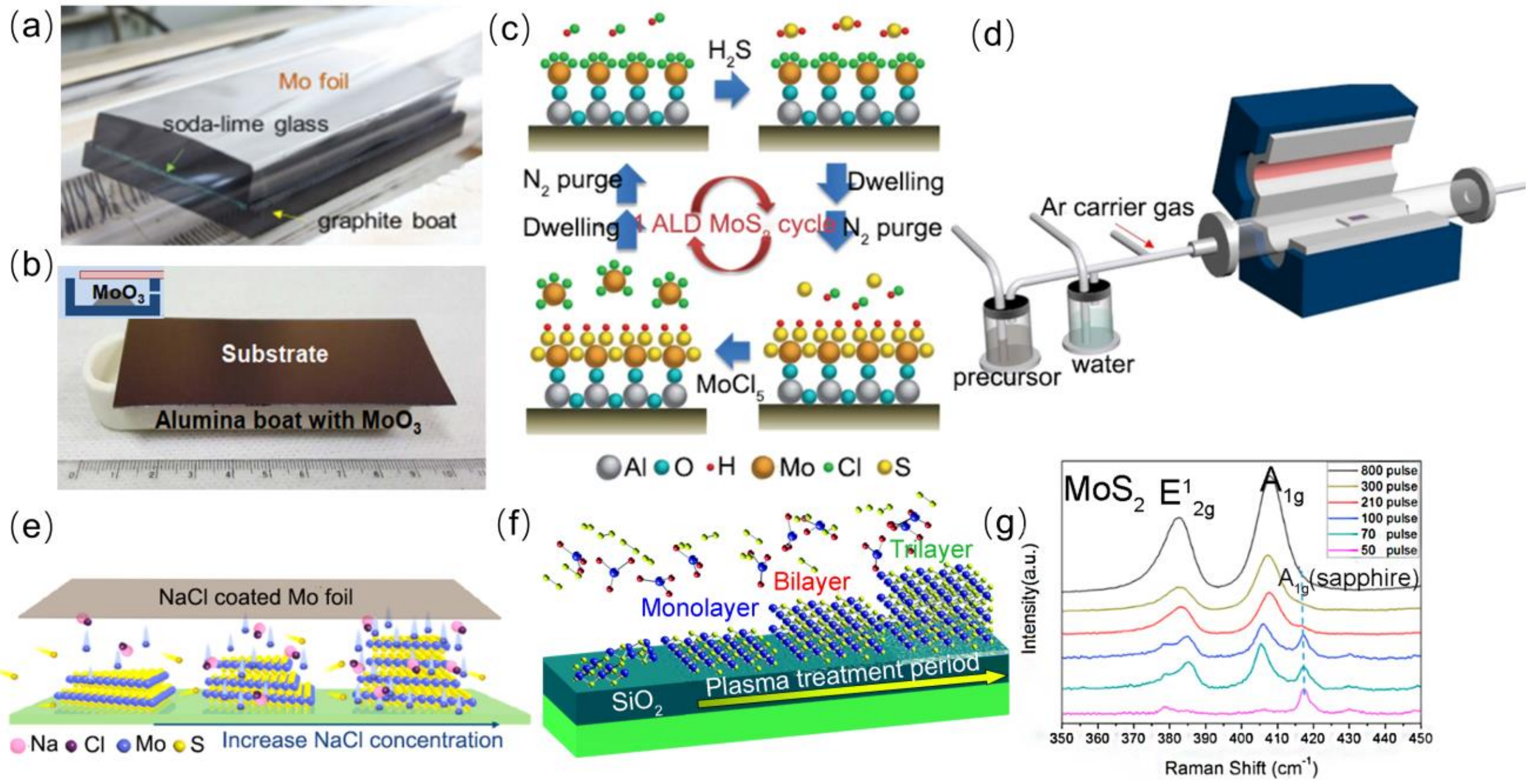

Figure 12. (a) Face-to-face configuration between the source and the substrate. (b) Point to source configuration. (c) Schematic showing growth of $MoS_2$ through reaction of ALD-deposited $MoCl_5$ and $H_2S$, and thickness control through cycles of ALD. (d) MOCVD Growth through a single mixture of $Mo(CO)_6$ , $C_2H_6S_2$, Ar/water vapor for thickness control. (e) $Na^+$-assisted growth with thickness control by concentration of NaCl solution. (f) Schematic showing the thickness increase with the oxygen plasma treatment period. (g) Raman spectra of the $MoS_2$ acquired by the pulse number in PLD. The distance between $E_{2g}$ and $A_{1g}$ indicates the variation of thickness. (a, d, e) Reproduced with permission from ref 111, 225. Copyright 2018, 2017, Springer Nature. (b, e) Reproduced with permission from ref 209, 226. Copyright 2017, 2019, American Chemical Society. (c, f) Reproduced with permission from ref 140, 227. Copyright 2014, 2015, Royal Society of Chemistry. (g) Reproduced with permission from ref 228. Copyright 2015, Wiley-VCH.

Gaseous phase precursor also provides better controllability than the solid precursor such as metal oxide and sulfur powder, resulting a better uniformity of the supply of precursors.[221-223] Wang et al. used $CS_2$ as sulfur source instead of sulfur powder for higher controllability to grow $WS_2$ and obtained larger domain size $WS_2$ with more regular shape.[224] Additionally, MOCVD, with the reactants delivered in gaseous phase, offers another route to supply the precursor uniformly. Choi et al. mixed molybdenum hexacarbonyl $Mo(CO)_6$ and dimethyl disulfide $C_2H_6S_2$ into an organic liquid precursor and used bubbler to drive the Ar/water vapor to carry out the gaseous phase precursor (Figure 12d). All the reactants mixed, allowing a uniform delivery of reactants and achieved 2cm-scale $MoS_2$ full coverage of different thickness. The growth time is a knob to control the thickness.[225] The wafer-scale $MoS_2$ film production has been increased to size of 15.24 cm (6-inch) recently.[112] Since the film grown by MOCVD is stitched by multiple domains, the issue of single-crystallinity still needs to be improved.

Furthermore, alkali-metal salt is also found to be a knob to adjust the thickness of the grown film. Yang et al., in another work, used NaCl coated Mo foil to grow $MoS_2$ on the soda-lime glass for thickness control (Figure 12e). They discover that, along with the increase of the concentration of NaCl solution for the immersion of the Mo foil, the layer number of the wedding-cake-like $MoS_2$ products could be tuned from ~2L to ~21L. Although the $MoS_2$ were discrete flakes, this study provides another hint for the layer number control of the film.[226]

Besides, Jeon et al discovered that the layer number could be adjusted by oxygen plasma treatment of the $SiO_2$ substrate prior to the growth. Varying the period (90s, 120s, or 300s) of the oxygen plasma treatment before the CVD can lead to fully covered mono-, bi- and tri-layers respectively (Figure 12f). However, the detailed mechanism of the variation of the numbers of layers due to plasma-treated $SiO_2$ substrates was not yet fully understood.[227]

Moreover, PLD offers another route to adjust the number of layers by controlling the number of laser pulses. Because of the relatively low temperature, reduced impurity due to the ultrahigh vacuum environment and the congruent transfer of the target species, the uniformity of the film growth is usually high. For instance, Ho et al. used the PLD to demonstrate the growth of high quality continuous $MoS_2$ thin films on the c-sapphire and controlled the layer number readily by manipulating the pulse number (Figure 12g). The in-situ RHEED can be used to monitor the thickness and surface cleanliness.[228]

Overall, in order to control the layer number of the grown film, more work is still needed to be devoted for developing a more uniform deposition condition. Beyond, clearer elucidation of the growth mechanism in the horizontal or vertical mode of the TMCs is also necessary.[229, 230]

## 3.3 Phase

TMCs have a unique property that they have a wide variety of phases.[231] The transition metal atoms are sandwiched by the chalcogen atoms. The three layer X(chalcogen)-M(metal)-X(chalcogen) can have different arrangements of ABA stacking or ABC stacking along z-axis. The ABA stacking forms a trigonal prismatic paradigm, called the 2H phase and the ABC stacking forms another octahedral paradigm, called the 1T phase (Figure 13a). In addition, a distorted octahedral phase (1T’) phase could be observed as well, especially for group 7 elements.[41, 206, 232] Likewise, some distortion of 2H phase will generate 2H’ phase. The structural distinctions transfer to the varied properties. 2H phase TMCs are semiconducting, while 1T phase TMCs are metallic. Other properties including mobility, thermal conductivity, Raman signatures etc. also varies.[233-235] As a result, controllability over phases offers the opportunity to manipulate the properties of the materials without adding or removing any atoms and combine different phase TMCs to fabricate novel devices. Thus phase control in the wafer-scale production is of great importance.[236]

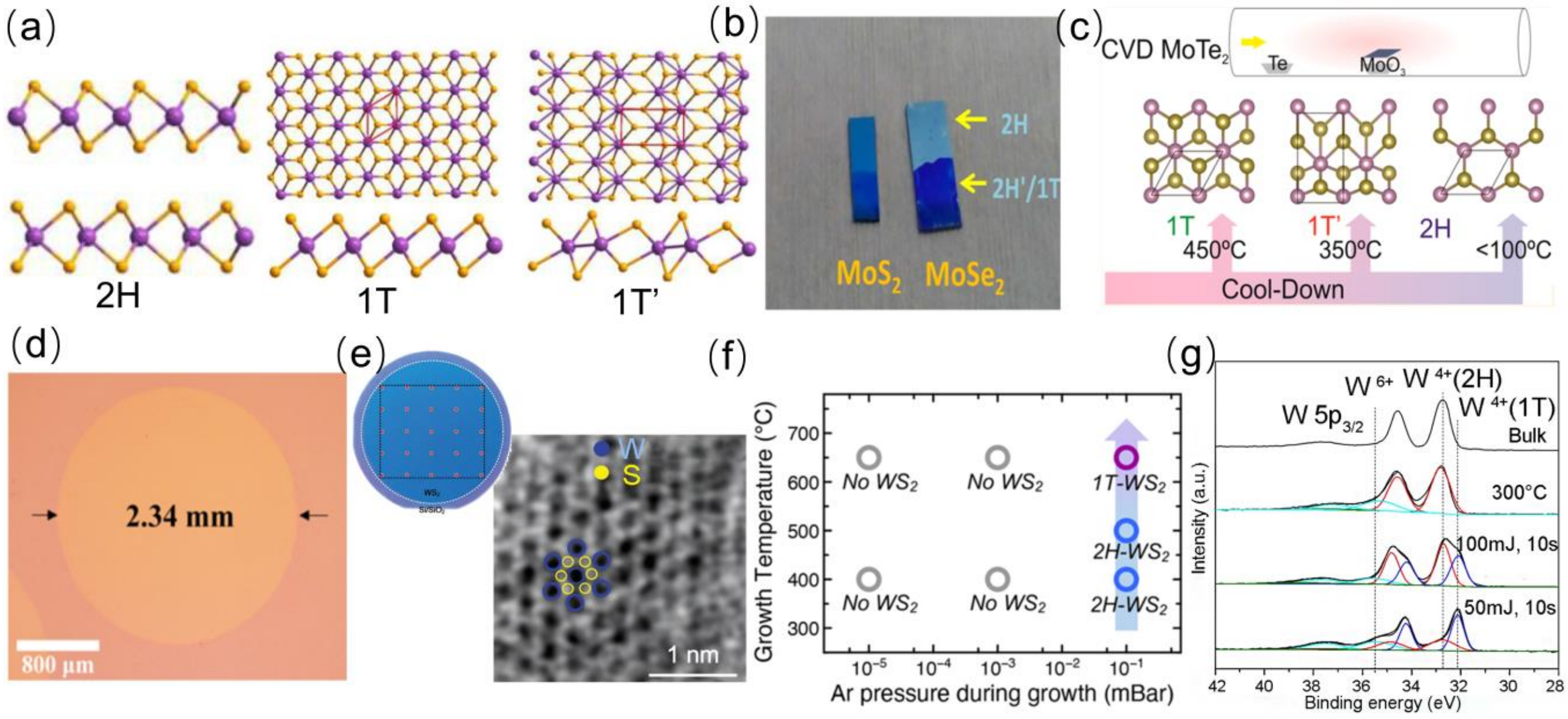

Figure 13. (a) Structural schematics of 2H, 1T, 1T’ phase TMCs. (b) $MoS_2$, $MoSe_2$ film partially converted to 2H’/1T phase with n-butyl lithium. (c) Synthesis of 1T, 1T’, 2H phase $MoTe_2$ by CVD with different temperature. (d) mm-scale conversion to 2H phase on 1T’phase $MoTe_2$. (e) 10.16 cm (4-inch) wafer scale production of 1T $WS_2$ with pre-deposited W film and $H_2S$ by PECVD. The TEM image features the 1T phase atom configuration. (f) Evolution of 2H-$WS_2$ to 1T-$WS_2$ which is resulted from the growth temperature of the substrate. (g) X-ray photoelectron spectra showing the increase of the signature of the 1T-$WS_2$. (a, e) Reproduced with permission from ref 232, 245. Copyright 2020, 2020, Wiley-VCH. (b, c, d, f) Reproduced with permission from ref 242, 243, 125, 167. Copyright 2016, 2017, 2019, 2020, American Chemical Society. (g) Reproduced from ref 247, under the terms of the Creative Commons Attribution 4.0 International License (http://creativecommons.org/licenses/by/4.0/). Copyright 2015, Springer Nature.

Generally, the phase obtained by the CVD approach is 2H phase and the 1T or 1T' phase is metastable.[81] Researchers have demonstrated to obtain metastable phase by methods such as electron beam irradiation,[237, 238] plasma treatment,[239] strain[240]. Kappera et al. discovered n-butyl lithium intercalation can convert 2H phase to 1T phase in the phase engineering. They displayed the device of 1T-2H-1T FET with an enhanced on current, mobility, on/off ratio etc. and lowered subthreshold swing, mainly due to diminished contact resistance.[241] Naz et al. used n-butyl-lithium to partially process thin film of $MoS_2$ and $MoSe_2$ to create a millimeter-scale region of 2H'/1T phase and form a heterostructure of continuous film (Figure 13b).[242] $MoTe_2$ is special in the aspect that the 1T' structure is more stable than the 2H phase in some range of temperature and Te deficiency.[125] Empante et al use the CVD to grow few layer $MoTe_2$ in the 2H, 1T' and 1T phases by different quenching temperature (Figure 13c).[243] However, these works are all about post-treatment. The scale is in most cases in micrometers and the homogeneity is not good.

Recently, some works about direct growth of the 1-T (T') phase film occurred.[234, 244] Xu et al. used magnetron sputtering to deposit a thin film of Mo and tellurized it to get few-layer metallic 1T'-$MoTe_2$ in wafer scale. It is found when growth time increases, the 1T'-$MoTe_2$ converts to 2H-$MoTe_2$. A 2.34 mm single-crystalline few-layer 2H-$MoTe_2$ was acquired (Figure 13d).[125] Moreover, Kim et al used the plasma-enhanced CVD (PECVD) to grow 1T $WS_2$ in a 10.16cm (4-inch) wafer scale with pre-deposited W film and $H_2S$ at a relatively low temperature 150 °C (Figure 13e).[245]

Besides, people found PLD approach can produce metastable phase directly. Wang et al. used the PLD to grow 0.25 $cm^2$ porous high-content 1T $MoS_2$ film and used a post sulfurization process to stabilize it.[246] Godel et al. found, with PLD deposition, under $10^{-1}$ mBar Ar pressure and 650 °C growth temperature, 1T phase began to appear on the 1×1cm $WS_2$ (Figure 13f).[167] Loh et al. also used PLD to deposit the $WS_2$ on a 1×1cm substrate and found a large portion of 1T $WS_2$ in the co-existence of 1T, 2H phases, under low laser energy and short deposition time (Figure 13g).[247] It is likely the energy from the laser pulse can induce phase transition. The mechanism has not been clarified yet.

The key for the phase conversion is to overcome the energy barrier, not very violently (or else it will do damage and bring about defects), and stabilize the material system.[234, 236] Overall, the phase-engineered large area TMCs growth is still in its infancy and more work about both the growth mechanism and the growth methodology is needed.

## 4. Device Applications of wafer-scale materials

The electronic and optoelectronic application of the TMCs largely depends on the wafer-scale production of the materials, in terms of both high quality and good spatial uniformity.[92, 248] Currently, many efforts have been devoted to develop electronic and optoelectronic devices on the wafer-scale TMCs.[80-82, 84, 249]

## 4.1 Electronics

Electronic application usually refers to devices including the field effect transistor (FET), p-n junctions, etc. Among them, FET is the most fundamental and significant component in the electronics. Several figure of merit are usually evaluated: (i) mobility (μ) indicates the responsive ability of the carriers to the applied electric field; (ii) on/off ratio ($I_{on}$/ $I_{off}$), which is equal to on-state current ($I_{on}$) over the off-state current ($I_{off}$), with high values standing for the effective switching and low power consumption; (iii) other parameters: subthreshold swing exhibits the transition speed between on and off states and Schottky barrier height indicates the energy barrier the carriers need to overcome between the working media and the electrical contacts.[81, 250-252]

The films grew by conventional powder method were used to fabricate electronic devices. Xu et al. used the traditional powder CVD with the Se powder and $PtCl_4$ powder to synthesized 2-layer poly-crystalline $PtSe_2$.[253] n-type or p-type behavior is directly correlated to the Se:Pt stoichiometric ratio. By tuning the amount of Se and $PtCl_4$, a continuous polycrystalline n- or p-type $PtSe_2$ film was attained in a 1×1cm scale. Arrays of field effect transistors (FETs) were fabricated and characterized. The current on/off ratio is about 25 for the n-type and 40 for the p-type devices. Annealing in the Ar/$H_2$ atmosphere for 2 h at 300°C reveals that field effect mobility $\mu_{FE}$ is enhanced to 10.8 and 10.43 $cm^2$ $V^{-1}$ $s^{-1}$ for the n-type and p-type 2-layer $PtSe_2$ (Figure 14a), suggesting annealing can reduce the surface traps and defects. Moreover, two p-type and n-type $PtSe_2$ FETs were connected by wire bonding to form a first complementary-metal-oxide-semiconductor (CMOS) inverter from CVD grown 2D-TMCs and successfully convert a logic "0" into a logic "1" and vice versa. Besides, an n-type film was transferred onto a p-type film, and p-n junction array was fabricated, exhibiting a typical rectifying behavior (Figure 14b). Limitations also exist: the on/off ratio for the CMOS ≈ 100, while CMOS application typically requires $10^5$-$10^6$.[253] Typical exfoliated single-crystalline $PtSe_2$ sheets can exhibit large on/off ratio of $10^3$-$10^5$ at room temperature. Improving crystal quality and domain size through further optimizing the $PtSe_2$ growth parameters, is key for an enhanced performance of the devices.

Electronic devices by the pre-deposited thin film sulfurization were also demonstrated: Chen et al. used the continuous film grown by placing a $SiO_2$/Si surface with a 5nm $WO_3$ thin film deposited by thermal evaporation above the substrate for growth, in a confined space of about 500 μm and then sulfurizing it. They fabricated a nearly 100 back-gate FET array in the wafer-scale and tested the properties (Figure 14c). The on/off ratio ranges from 10 to $10^5$ and the mobility is in the range of 0.2 - 4.0 $cm^2\ V^{-1}\ s^{-1}$, with the most probable value of around 1.3 $cm^2\ V^{-1}\ s^{-1}$, smaller than the reported best result of 18 $cm^2\ V^{-1}\ s^{-1}$ by Kang's seeding paper.[113, 211] This relatively low mobility is likely to be resulted from the prescence of grain boundaries in the randomly defined channels in FET arrays. Better performance still fundamentally rely on the quality of the film, especially the improved grain boundaries and uniformity.

Song et al pre-deposit W film on 10.16cm (4-inch) $SiO_2$/Si substrate and followed by powder-based tellurization to fabricate wafer-size $WTe_2$ film. They also demonstrated wafer-scale position-controlled $WTe_2$ pattern with the W shapes defined beforehand by photolithography and a metal reactive ion etcher. Then they transferred CVD-grown $MoS_2$ film with wet transfer to the as-grown $WTe_2$ patterns and fabricated the $MoS_2/WTe_2$ heterostructures (Figure 14d). The electrical characterization reveals an average Schottky barrier height (SBH) of the $WTe_2/MoS_2$ interface as ~98.9 meV (Figure 14e), very close to the theoretical Schottky-Mott limit value (~100 meV).[254] It implied the high quality material can be prepared by the pre-deposition approach.

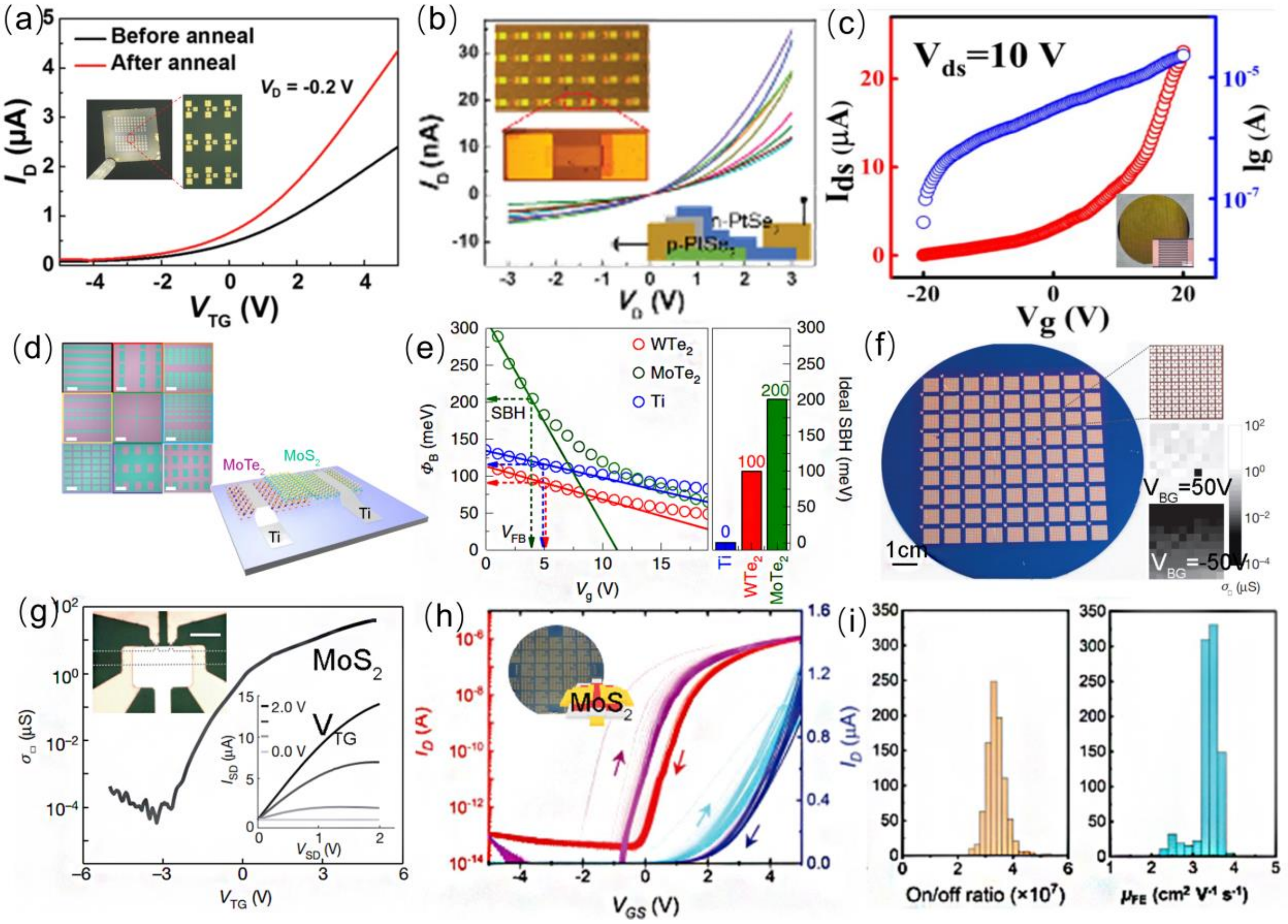

Figure 14. (a) Transfer characteristic of n-$PtSe_2$ FET. (b) Rectifying behaviors of the arrays of p-n junction of $PtSe_2$. (c) Transfer characteristic of the $WS_2$ FET. (d) Patterns of the $WTe_2$ and $WTe_2/MoS_2$ heterostructure by tellurization of pre-deposited W film. (e) The Schottky barrier height of the $MoS_2/WTe_2$ and $MoS_2/MoTe_2$ interfaces. (f) Arrays of FETs fabricated on the $MoS_2$ grown by MOCVD. (g) Transfer characteristic of a typical FET in f. Inset is the output characteristic. (h) Transfer characteristics of the 900 FETs on the 15.24 cm (6-inch) $MoS_2$. (i) Histograms of n/off ratios and field effect mobilities. (a, b, c) Reproduced with permission from ref 253, 211. Copyright 2019, American Chemical Society. (d, e, f, g, h, i) Reproduced with permission from ref 254, 113, 112. Copyright 2020, 2015, 2020 Springer Nature.

Moreover, in the seeding work to employ MOCVD to grow wafer-scale continuous TMCs film, Kang et al.  grew continuous and homogeneous $MoS_2$ and $WS_2$ film and batch-fabricated 8,100 $MoS_2$ FET devices with a global back gate, via the standard photolithography (Figure. 14f). They showed a high on/off conductance ratio (~$10^6$), a relatively high field effect mobility (~29 $cm^2\ V^{-1}\ s^{-1}$) and large transconductance (~2 $\mu S\ \mu m^{-1}$) (Figure 14g). More practically, they deposited a 500nm thick layer of $SiO_2$, grew the second monolayer $MoS_2$, and then fabricated the second layer of FET arrays.[113] Both arrays show similar field-effect mobilities and transfer characteristics, indicating the feasibility to develop a three-dimensional device architectures based on the MOCVD-grown TMCs film.

Furthermore, Seol et al. demonstrated high-throughput growth of monolayer $MoS_2$ film with maximum size of 15.24 cm (6-inch) by MOCVD with high spatial uniformity. They used the wafer-scale film to fabricate 900 field effect transistors (FET) in the same wafer and the electrical characterization of all the 900 FETs showed excellent uniformity of high on-current of $1.1 \pm 0.1$ μA (Figure 14h), extremely low off-current of $32.0 \pm 3$ fA, high on/off ratio ($10^7$) and field effect mobility (3.4 $cm^2 \cdot V^{-1} \cdot s^{-1}$) (Figure 14i).[112] It points to a route of integrating the 2D TMCs to the current Si CMOS platforms.

The performance of the wafer-scale devices based on 2D TMCs still roots in the quality of the materials, i.e. less grain boundaries, greater grain size, and better uniformity of the layer number. For example, in the MOCVD growth, Figure 6c shows the randomly aligned grain boundaries in the poly-crystals. It can significantly lower the uniformity of the device applications. Some strategies for improvement have been partly discussed in the section for the optimization of the grains. This calls for a further advance in the methodology and theoretical understandings of the 2D TMCs growth. Last but not least, integration of the 2D TMCs to the current mature fabrication art of the semiconductors seamlessly needs more investigation.

## 4.2 Optoelectronics

An optoelectronic device features a photon-to-charge conversion or vice versa. Typical applications include photodetectors, light emitting device (LED), and photovoltaics. The most devices fabricated for the wafer-scale 2D TMCs so far are photodetectors.[81] The parameters of responsivity (R), which equals to the difference value of the on and off current over the incident power, and also the on/off ratio are frequently evaluated.[250, 255]

Hoang et al. grew a layer of graphene onto 5.08cm (2-inch) c-axis sapphire (0001) substrate and epitaxialy grew $MoS_2$ film in a horizontal, hot wall MOCVD system. They fabricated 900 $MoS_2$/graphene heterostructure photo-detectors (PD). When stimulated with light, the net hole concentration in the p-type graphene was reduced by the injected electrons from $MoS_2$, resulting in a decrease in conductivity, and thus the current dropped under light illumination. When the illumination was discontinued, the current recovered, reversed to the traditional photo-detector (Figure 15a, b). The histogram of the on/off current ratio ($I_{on}/I_{off}$) exhibits an average on/off ratio as ~$2.5^{-1}$, with ultrahigh photo-responsivity ~43426 $A \cdot W^{-1}$ at wavelength of 475nm. After 100 cycles, the PDs still showed good cycling stability and reversibility.[256]

Zeng et al. pre-deposited palladium film and selenized it with the selenium powder to grow ~50-layer 1.5×1.5 cm continuous $PdSe_2$ film. 4×4 array of PDs. Solution-processable black phosphorus quantum dots (BPQDs) was decorated onto top surface of the $PdSe_2$ film by spin-coating method to optimize the device performance (Figure 15c). Stimulated by the 780nm laser, a responsivity (R) of 300.2 $mA \cdot W^{-1}$ and a detectivity ($D^*$) of $1.18\times10^{13}$ Jones were obtained at low light intensity of 0.9 $\mu W \cdot cm^{-2}$ at 0 V, comparable with the responsivity of graphene/Si Schottky junction PD (270-730 $mA \cdot W^{-1}$),[257-259] greater than that of $PtSe_2$ based PDs (10-262 $mA \cdot W^{-1}$).[260-262] The relatively quick response speed of 38/44 μs (rise/fall time) for BPQDs @ $PdSe_2$/Si were determined experimentally (Figure 15d), faster than the graphene/Si Schottky junction (93/110 μs).[263] The $PdSe_2$ film, as building blocks, potentially helps to assemble next generation optoelectronic system.[264]

Han et al. grew wafer-scale $MoS_2$ and $PtTe_2$ continuous film on NaCl substrate by sulfurizing/tellurizing pre-deposited Mo or Pt films and integrated the 2D $PtTe_2$ layer on two ends of 2D $MoS_2$ layer on a flexible polyethylene terephthalate (PET) substrate to create flexible photo-detection device. The intrinsically metallic nature of $PtTe_2$ layer (extremely high electrical conductivities, $>10^6$ S/m) enabled atom-thickness metallic electrodes for the semiconducting 2D $MoS_2$ layers. A periodic generation of photocurrent upon a cyclic illumination of visible light could be realized (Figure 15e). This confirm a concept of large area "all 2D" photo-responsive device.[129] From the photo-response curve, the on-off ratio still needs improvement.

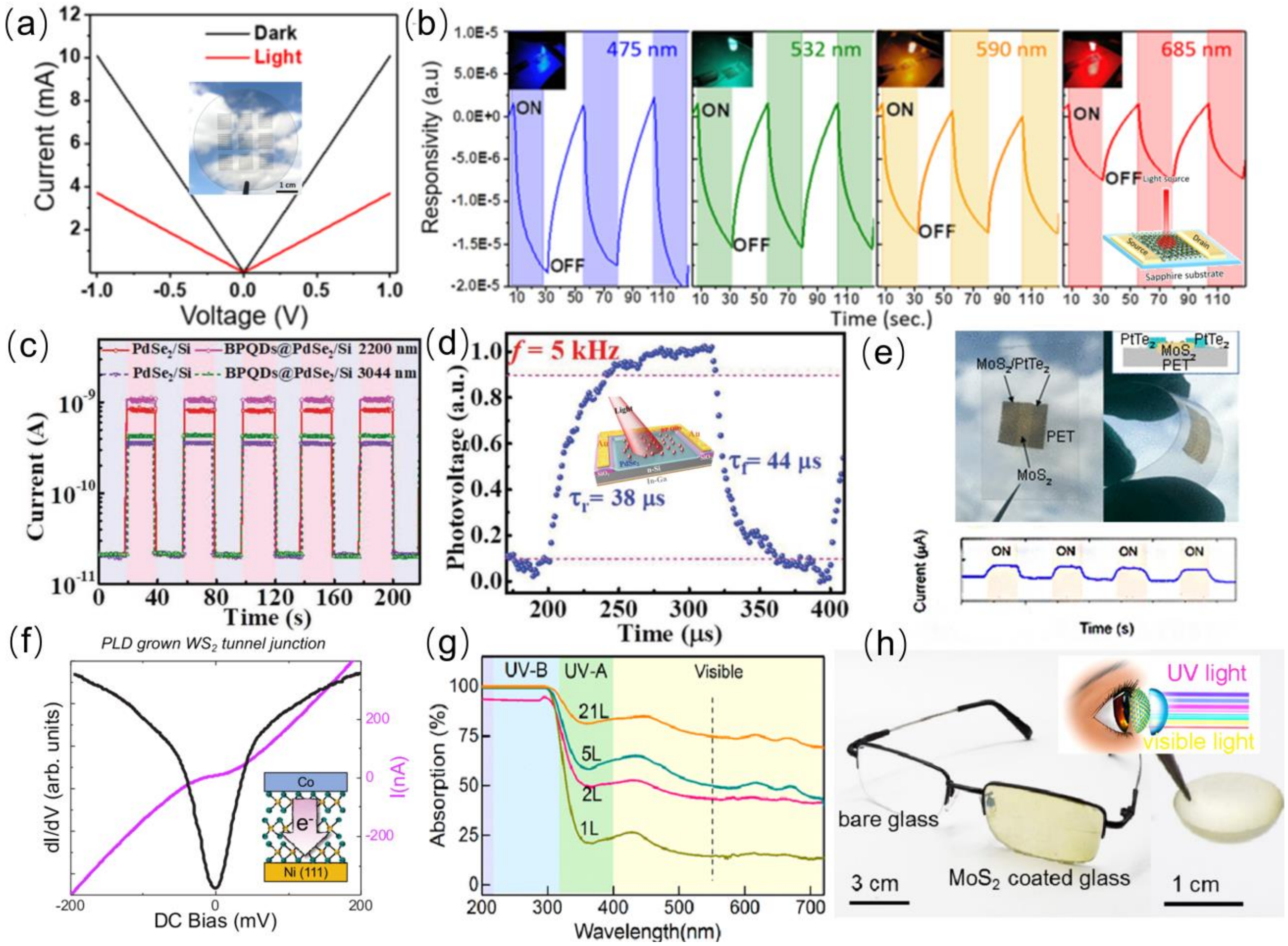

Figure 15. (a) Light and dark current of the $MoS_2$/graphene PD. Inset shows the PD arrays. (b) Wavelength dependent photo-response with insets showing the device cartoon. (c) Temporal photo-response of $PdSe_2$/Si with BPQDs decoration under 2200, 3044nm illumination. (d) A single normalized cycle of the BPQDs@$PdSe_2$/Si PD for response speed. (e) An "all-2D" $PtTe_2$/$MoS_2$ PD on flexible PET substrate and the photo-response. (f) I(V) and dI/dV characterizations of Ni/$WS_2$/Co tunneling junction. (g) UV-vis absorption spectra of different $MoS_2$ layers. (h) Prototype UV-absorbing $MoS_2$-coated spectacle and contact lens. (a, b, e, f, g, h) Reproduced with permission from ref 256, 129, 167, 226. Copyright 2020, 2020, 2020, 2019, American Chemical Society. (c, d) Reproduced with permission from ref 264. Copyright 2019, Wiley-VCH.

On-off ratio of the current photodetectors based on TMCs like MoS are generally still not high enough. The cause can be in many aspects： the crystallinity is not ideal; the yet low photo-carrier generation of $MoS_2$ due to that monolayer thickness limits the absorption of incident light; the contact barrier of the TMCs and electrodes. The most fundamental issue is to increase the quality of the grown material. Besides, combining with highly photosensitve material such as quantum dots is very conducive to enhance photosensitivity, especially with the consideration that 2D TMC itself is an interface. The studies of other types of optoelectronics such as LED and photovoltaics in wafer-scale TMCs are still not abundant and more investigations are necessary.

## 4.3 Other types of devices

While the mainstream applications of the 2D TMCs film are electronic and optoelectronic devices. Some other types of applications were also studied.

Godel et al. used the PLD to grow wafer scale $WS_2$ ultrathin layers and fabricated a tunneling hetero-structure with $WS_2$ as a tunneling barrier. Co and Ni(111) were used for the two vertical electrical contacts, allowing one to probe the vertical tunneling transport through the nearly insulating $WS_2$ layers. The current I(V) and its first derivative dI/dV at 4K exhibit the typical characteristics of tunnel junctions with good reproducibility, displaying the compatibility of the wafer-size TMCs film with vertical device integration (Figure 15f).[167] It highlights the possibility to integrate TMCs material into the spintronic applications.

Moreover, Yang et al. spun coat NaCl solution onto the Mo foil as Mo precursor and then grew $MoS_2$ on glass substrates. They exploited the properties of the $MoS_2$ with high absorption of the UV light section (beyond 95% for the UV-B section) but sufficient transmission in the visible light absorption to grow the continuous $MoS_2$ film onto the glass, in order to manufacture UV blocking sunglass (Figure 15g). The transmission of the section of visible light can be tuned by the variable thicknesses of the $MoS_2$ film coating, affording different levels of protection. They fabricated prototype UV-filtering rigid spectacles and soft contact lenses (Figure 15h), utilizing the flexibility, hydrophilicity, biocompatibility of the atomically thin $MoS_2$.[111]

Since the production of wafer-scale 2D TMCs is still challenging, currently there are limited works for the applications of the wafer-scale TMCs. The device performance depends on the quality and homogeneity of the grown materials, and thus calls for more advances for the high-quality wafer-scale TMCs. Moreover, the inversion symmetry breaking and strong spin-orbit interactions result in properties such as spin hall effect[265, 266] and two energetic degenerate inequivalent valleys,[69] making 2D TMCs promising in the non-linear optics,[267] spintronics and valleytronics.[268] However, the corresponding device applications, especially in the wafer-scale are still deficient. Besides, the realm of wafer-scale devices based on phase-engineered TMCs is also unvisited. There is still plenty of room for the exploration.

## 5. Concluding remarks and outlook

In summary, 2D TMCs, with plethora of superior properties, making itself extremely promising for the fundamental physical issues and high performance applications in a variety of devices.[51, 249] The last around five years saw the large enhancement of the TMC growth size for different approaches, from the scale of millimeter to decimeter.[84] Prominently, salt assisted growth, MOCVD, approaches were reported to reach maximum growth size of 1.5 dm (6-inch). [111, 112] Herein, we first reviewed the advances recent progress brings about, including different approaches to achieve wafer-scale 2D TMCs growth. The advantages and disadvantages of each approaches are illuminated. Furthermore, current challenges are the enhancement of the grain size (optimal to be large single-crystalline wafer), control on the morphology, layer number and phase. We further analyze these factors and optimization for the wafer-scale 2D TMCs growth. In addition, we stated various applications of the wafer-scale TMCs, including the mainstream electronic and optoelectronic devices, and also other types of applications. Here we also envision some aspects for particular attention to further the advancement of wafer-scale production of 2D TMCs (1-5 are about setup and growth conditions, 6-8 are about material quality and variety, 9 is about expansion in the aspect of application):

(1) Currently, the typical technique to grow wafer-scale TMCs is CVD in the tube furnace. In the tube furnace, many factors such as the evenness of the gas flow, the directional uniformity of the evaporation of the solid precursors, the thermal field, the local motion of flow near the substrate surface, altogether influence the growth results, in an entangled manner. Whether we can do some innovations or adaptions to the CVD system, for example, creating an evener gas flow, could lead to an improved growth result. Tang's work to use vertical CVD configuration is a good trial in this aspect.[148] The situation is the same as for other setups, such as MOCVD, PLD etc.

(2) As for the reactants, how to make the distribution evener is an extremely interesting research question. For the precursors, a “face to face” configuration to the substrate is superior to the “point to face” configuration. But more possibilities can be explored. For the salt-assisted growth, research questions, for example, the distribution and motion of the alkali metal atoms in the substrate such as soda-lime glass during the growth are worth more investigations.

(3) The substrate is also a significant issue for growth. Si maybe is not the best candidate for growth of 2D TMCs, due to the symmetry not matching well.[197, 269] The TMCs do not grow readily on Si in general.[270] Will other substrates like GaN, 4H- or 6H-SiC be good candidates for wafer-scale growth? They have hexagonal symmetry, the same with many TMCs. making it easier to control the alignment of the grains. In addition, they are also significant semiconductors in current industry. The epitaxy of TMCs on these substrates and corresponding devices are worthy of research.

(4) For the current growth with the pre-deposited film, MOCVD or conventional powder CVD, the growth duration can last for tens of minutes, to even 20 hours, without the ramping and cooling time taken into account of.[113] Thus fast growth of the wafer scale 2D TMCs is especially desired.

(5) For growth of $MoS_2$ or $WS_2$, it calls for temperature above 500℃. For flexible device applications, it is optimal to grow the TMCs on the flexible substrate directly.[271] The polymeric substrate usually cannot endure high temperature.[272] Thus, development of low temperature growth of wafer-scale TMCs is necessary and anticipated.

(6) As we know, larger domain size will lead to less grain boundaries, providing a route to enhance the materials to higher quality and better performance. The most ideal case is the single-crystalline large size TMCs. Recently, similar to the single-crystalline Si growth, growth of graphene by using a local feed setup, starting from a single nuclei, was reported.[273] Using the single nucleus method is challenging for TMCs growth due to the binary/ ternary composition and frequently used solid precursors. Using metal-organic precursors or liquid precursors could also be a possible solution. It calls for more exploration.

(7) Currently, the wafer-scale TMCs growth is largely about group VIB $MoS_2$ and $WS_2$. Studies of other wafer-scale 2D TMCs, for example, $NbSe_2$, $Bi_2Se_3$, $Bi_2O_2Se$ will be novel. The wafer-scale alloy TMCs like $WSe_{2-x}S_x$ or "Janus" TMCs like Mo-S-Se are also worthy of study.

(8) Heterostructure combines more than one material and generates novel properties beyond each single one.[274] Twist angle further adds another degree of freedom and induced intriguing properties.[275] Although state-of-art vertical stacking technique can fabricate heterostructure, it is multiple-step aligning transfer procedure, thus time consuming and not applicable to large-scale production. The research about direct growth of wafer-scale heterostructure is awaiting development.

(9) Finally, in addition to the electronic and optoelectronic properties such as descent mobilities and bandgap tunability, TMCs still possess other unique properties: for example, inversion symmetry breaking, strong spin-orbit interactions and valley properties, resulting in promising opportunities in non-linear optics, spintronics and valleytronics. The previous works about the device applications of the wafer-scale TMCs are still not abundant and concentrate in electronics and optoelectronics. Besides, the territory of devices based on wafer-scale phase-engineered TMCs is not yet reclaimed. Thus, researchers should pioneer to explore other device structures and exploit more in this rich mine.

**Conflict of Interest**

The authors declare no conflict of interest.

**Author Information**

Corresponding Authors:

*E-mail: pjwangpku@zju.edu.cn (P.W.)

*E-mail: xdpi@zju.edu.cn (X.P.)

**Acknowledgements**

We acknowledge support from National Natural Science Foundation of China (Grant Nos. 51902061, 62090031 and 91964107).

# References

[1] S. E. Root, S. Savagatrup, A. D. Printz, D. Rodriquez, D. J. Lipomi, *Chem. Rev.* **2017**, *117*, 6467.

[2] C. E. Leiserson, N. C. Thompson, J. S. Emer, B. C. Kuszmaul, B. W. Lampson, D. Sanchez, T. B. Schardl, *Science* **2020**, *368*, 1079.

[3] M. Dragoman, A. Dinescu, D. Dragoman, *Phys. Status Solidi (a)* **2019**, *216*, 1800724.

[4] C. Liu, L. Wang, J. Qi, K. Liu, *Adv. Mater.* **2020**, *32*, 2000046.

[5] R. Zhang, T. Chen, A. Bunting, R. Cheung, *Microelectron. Eng.* **2016**, *154*, 62.

[6] M.-Y. Li, *Nature* **2019**, *567*, 169.

[7] A. D. Franklin, *Science* **2015**, *349*, 2750.

[8] J. Fang, Z. Zhou, M. Xiao, Z. Lou, Z. Wei, G. Shen, *InfoMat* **2019**, *2*, 291.

[9] F. F. Wang, X. Y. Hu, X. X. Niu, J. Y. Xie, S. S. Chu, Q. H. Gong, *Journal of Materials Chemistry C* **2018**, *6*, 924.

[10] L. Men, M. A. White, H. Andaraarachchi, B. A. Rosales, J. Vela, *Chem. Mater.* **2016**, *29*, 168.

[11] D. Zhou, P. Wang, C. R. Roy, M. D. Barnes, K. R. Kittilstved, *J. Phys. Chem. C* **2018**, *122*, 18596.

[12] J. M. Pietryga, Y. S. Park, J. Lim, A. F. Fidler, W. K. Bae, S. Brovelli, V. I. Klimov, *Chem. Rev.* **2016**, *116*, 10513.

[13] Y. Zhu, W. Huang, Y. He, L. Yin, Y. Zhang, D. Yang, X. Pi, *Research* **2020**, *2020*, 7538450.

[14] Z. Ni, S. Zhou, S. Zhao, W. Peng, D. Yang, X. Pi, *Mater. Sci. Eng. R Rep.* **2019**, *138*, 85.

[15] V. Schmidt, J. V. Wittemann, S. Senz, U. Gösele, *Adv. Mater.* **2009**, *21*, 2681.

[16] E. Garnett, L. Mai, P. Yang, *Chem. Rev.* **2019**, *119*, 8955.

[17] C. Jia, Z. Lin, Y. Huang, X. Duan, *Chem. Rev.* **2019**, *119*, 9074.

[18] A. M. Morales, C. M. Lieber, *Science* **1998**, *279*, 208.

[19] Y. Cui, Q. Wei, H. Park, C. M. Lieber, *Science* **2001**, *293*, 1289.

[20] S. Park, M. Vosguerichian, Z. Bao, *Nanoscale* **2013**, *5*, 1727.

[21] V. Schroeder, S. Savagatrup, M. He, S. Lin, T. M. Swager, *Chem. Rev.* **2019**, *119*, 599.

[22] R. Rao, C. L. Pint, A. E. Islam, R. S. Weatherup, S. Hofmann, E. R. Meshot, F. Wu, C. Zhou, N. Dee, P. B. Amama, J. Carpena-Nunez, W. Shi, D. L. Plata, E. S. Penev, B. I. Yakobson, P. B. Balbuena, C. Bichara, D. N. Futaba, S. Noda, H. Shin, K. S. Kim, B. Simard, F. Mirri, M. Pasquali, F. Fornasiero, E. I. Kauppinen, M. Arnold, B. A. Cola, P. Nikolaev, S. Arepalli, H. M. Cheng, D. N. Zakharov, E. A. Stach, J. Zhang, F. Wei, M. Terrones, D. B. Geohegan, B. Maruyama, S. Maruyama, Y. Li, W. W. Adams, A. J. Hart, *ACS Nano* **2018**, *12*, 11756.

[23] L. Chen, H. He, S. Zhang, C. Xu, J. Zhao, S. Zhao, Y. Mi, D. Yang, *Nanoscale Res Lett* **2013**, *8*, 225.

[24] N. Du, H. Zhang, X. Ma, D. Yang, *Chem. Commun.* **2008**, 6182.

[25] K. S. Novoselov, A. K. Geim, S. V. Morozov, D. Jiang, Y. Zhang, S. V. Dubonos, I. V. Grigorieva, A. A. Firsov, *Science* **2004**, *306*, 666.

[26] K. I. Bolotin, in *Graphene*, 2014, 199.

[27] L. Banszerus, M. Schmitz, S. Engels, J. Dauber, M. Oellers, F. Haupt, K. Watanabe, T. Taniguchi, B. Beschoten, C. Stampfer, *Sci Adv* **2015**, *1*, e1500222.

[28] X. Duan, C. Wang, A. Pan, R. Yu, X. Duan, *Chem. Soc. Rev.* **2015**, *44*, 8859.

[29] S. Manzeli, D. Ovchinnikov, D. Pasquier, O. V. Yazyev, A. Kis, *Nat. Rev. Mater.* **2017**, *2*, 17033.

[30] T. Chowdhury, E. C. Sadler, T. J. Kempa, *Chem. Rev.* **2020**, *120*, 12563.

[31] J. D. Caldwell, I. Aharonovich, G. Cassabois, J. H. Edgar, B. Gil, D. N. Basov, *Nat. Rev. Mater.* **2019**, *4*, 552.
[32] K. Zhang, Y. Feng, F. Wang, Z. Yang, J. Wang, *Journal of Materials Chemistry C* **2017**, *5*, 11992.
[33] F. Xia, H. Wang, J. C. M. Hwang, A. H. C. Neto, L. Yang, *Nat. Rev. Phys.* **2019**, *1*, 306.
[34] X. Ling, H. Wang, S. Huang, F. Xia, M. S. Dresselhaus, *Proc. Natl. Acad. Sci. U. S. A.* **2015**, *112*, 4523.
[35] H. Oughaddou, H. Enriquez, M. R. Tchalala, H. Yildirim, A. J. Mayne, A. Bendounan, G. Dujardin, M. Ait Ali, A. Kara, *Prog. Surf. Sci.* **2015**, *90*, 46.
[36] Z. Shi, R. Cao, K. Khan, A. K. Tareen, X. Liu, W. Liang, Y. Zhang, C. Ma, Z. Guo, X. Luo, H. Zhang, *Nano-Micro Lett.* **2020**, *12*, 99.
[37] B. Liu, M. Kopf, A. N. Abbas, X. Wang, Q. Guo, Y. Jia, F. Xia, R. Weihrich, F. Bachhuber, F. Pielnhofer, H. Wang, R. Dhall, S. B. Cronin, M. Ge, X. Fang, T. Nilges, C. Zhou, *Adv. Mater.* **2015**, *27*, 4423.
[38] E. Gibney, *Nature* **2015**, *522*, 274.
[39] Q. Yun, L. Li, Z. Hu, Q. Lu, B. Chen, H. Zhang, *Adv. Mater.* **2020**, *32*, e1903826.
[40] D. Jariwala, T. J. Marks, M. C. Hersam, *Nat. Mater.* **2017**, *16*, 170.
[41] J. You, M. D. Hossain, Z. Luo, *Nano Converg.* **2018**, *5:26*, 1.
[42] D. L. Duong, S. J. Yun, Y. H. Lee, *ACS Nano* **2017**, *11*, 11803.
[43] H. Tian, M. L. Chin, S. Najmaei, Q. Guo, F. Xia, H. Wang, M. Dubey, *Nano Res.* **2016**, *9*, 1543.
[44] A. A. Tedstone, D. J. Lewis, P. O'Brien, *Chem. Mater.* **2016**, *28*, 1965.
[45] M. Samadi, N. Sarikhani, M. Zirak, H. Zhang, H. L. Zhang, A. Z. Moshfegh, *Nanoscale Horiz.* **2018**, *3*, 90.
[46] L. Pi, L. Li, K. Liu, Q. Zhang, H. Li, T. Zhai, *Adv. Funct. Mater.* **2019**, *29*, 1904932.
[47] M. Pumera, Z. Sofer, A. Ambrosi, *J. Mater. Chem. A* **2014**, *2*, 8981.
[48] M. J. Mleczko, C. Zhang, H. R. Lee, H. H. Kuo, B. Magyari-Kope, R. Moore, Z. X. Shen, I. R. Fisher, Y. Nishi, E. Pop, *Sci. Adv.* **2017**, *3*, e1700481.
[49] T. Kanazawa, T. Amemiya, A. Ishikawa, V. Upadhyaya, K. Tsuruta, T. Tanaka, Y. Miyamoto, *Sci. Rep.* **2016**, *6*, 22277.
[50] G. Li, X. Wang, B. Han, W. Zhang, S. Qi, Y. Zhang, J. Qiu, P. Gao, S. Guo, R. Long, Z. Tan, X. Z. Song, N. Liu, *J. Phys. Chem. Lett.* **2020**, *11*, 1570.
[51] Y. Zhang, Y. Yao, M. G. Sendeku, L. Yin, X. Zhan, F. Wang, Z. Wang, J. He, *Adv. Mater.* **2019**, 1901694.
[52] A. Zhang, X. Ma, C. Liu, R. Lou, Y. Wang, Q. Yu, Y. Wang, T.-l. Xia, S. Wang, L. Zhang, X. Wang, C. Chen, Q. Zhang, *Phys. Rev. B* **2019**, *100*, 201107.
[53] J. C. Rasch, T. Stemmler, B. Muller, L. Dudy, R. Manzke, *Phys. Rev. Lett.* **2008**, *101*, 237602.
[54] R. Lv, J. A. Robinson, R. E. Schaak, D. Sun, Y. Sun, T. E. Mallouk, M. Terrones, *Acc. Chem. Res.* **2015**, *48*, 56.
[55] H. Wang, X. Huang, J. Lin, J. Cui, Y. Chen, C. Zhu, F. Liu, Q. Zeng, J. Zhou, P. Yu, X. Wang, H. He, S. H. Tsang, W. Gao, K. Suenaga, F. Ma, C. Yang, L. Lu, T. Yu, E. H. T. Teo, G. Liu, Z. Liu, *Nat. Commun.* **2017**, *8*, 394.
[56] P. M. Coelho, K. Nguyen Cong, M. Bonilla, S. Kolekar, M.-H. Phan, J. Avila, M. C. Asensio, I. I. Oleynik, M. Batzill, *J. Phys. Chem. C* **2019**, *123*, 14089.
[57] A. Jalouli, M. Kilinc, P. Wang, H. Zeng, T. Thomay, *J. Appl. Phys.* **2020**, *128*, 124304.
[58] L. M. Xie, *Nanoscale* **2015**, *7*, 18392.
[59] X. Duan, C. Wang, Z. Fan, G. Hao, L. Kou, U. Halim, H. Li, X. Wu, Y. Wang, J. Jiang, A. Pan, Y. Huang, R. Yu, X. Duan, *Nano Lett.* **2016**, *16*, 264.

[60] A. Splendiani, L. Sun, Y. Zhang, T. Li, J. Kim, C. Y. Chim, G. Galli, F. Wang, *Nano Lett.* **2010**, *10*, 1271.
[61] G. Eda, H. Yamaguchi, D. Voiry, T. Fujita, M. Chen, M. Chhowalla, *Nano Lett.* **2011**, *11*, 5111.
[62] M. Kang, B. Kim, S. H. Ryu, S. W. Jung, J. Kim, L. Moreschini, C. Jozwiak, E. Rotenberg, A. Bostwick, K. S. Kim, *Nano Lett.* **2017**, *17*, 1610.
[63] J. R. Schaibley, H. Yu, G. Clark, P. Rivera, J. S. Ross, K. L. Seyler, W. Yao, X. Xu, *Nat. Rev. Mater.* **2016**, *1*, 16055.
[64] K. F. Mak, K. He, J. Shan, T. F. Heinz, *Nat. Nanotechnol.* **2012**, *7*, 494.
[65] A. Manchon, H. C. Koo, J. Nitta, S. M. Frolov, R. A. Duine, *Nat. Mater.* **2015**, *14*, 871.
[66] H. Li, S. Ruan, Y. J. Zeng, *Adv. Mater.* **2019**, *31*, 1900065.
[67] Z. Wang, J. Shan, K. F. Mak, *Nat. Nanotechnol.* **2017**, *12*, 144.
[68] L. Guo, M. Wu, T. Cao, D. M. Monahan, Y.-H. Lee, S. G. Louie, G. R. Fleming, *Nat. Phys.* **2018**, *15*, 228.
[69] K. F. Mak, K. L. McGill, J. Park, P. L. McEuen, *Science* **2014**, *344*, 1489.
[70] H. Yu, X. Cui, X. Xu, W. Yao, *Natl. Sci. Rev.* **2015**, *2*, 57.
[71] Y. C. Zou, Z. G. Chen, E. Zhang, F. Xiu, S. Matsumura, L. Yang, M. Hong, J. Zou, *Nanoscale* **2017**, *9*, 16591.
[72] D. Wickramaratne, S. Khmelevskyi, D. F. Agterberg, I. I. Mazin, *Phys. Rev. X* **2020**, *10*, 041003
[73] W. Yu, J. Li, T. S. Herng, Z. Wang, X. Zhao, X. Chi, W. Fu, I. Abdelwahab, J. Zhou, J. Dan, Z. Chen, Z. Chen, Z. Li, J. Lu, S. J. Pennycook, Y. P. Feng, J. Ding, K. P. Loh, *Adv. Mater.* **2019**, *31*, 1903779.
[74] A. O. Fumega, M. Gobbi, P. Dreher, W. Wan, C. González-Orellana, M. Peña-Díaz, C. Rogero, J. Herrero-Martín, P. Gargiani, M. Ilyn, M. M. Ugeda, V. Pardo, S. Blanco-Canosa, *J. Phys. Chem. C* **2019**, *123*, 27802.
[75] G. Duvjir, B. K. Choi, I. Jang, S. Ulstrup, S. Kang, T. Thi Ly, S. Kim, Y. H. Choi, C. Jozwiak, A. Bostwick, E. Rotenberg, J. G. Park, R. Sankar, K. S. Kim, J. Kim, Y. J. Chang, *Nano Lett.* **2018**, *18*, 5432.
[76] J. Kuneš, L. Baldassarre, B. Schächner, K. Rabia, C. A. Kuntscher, D. M. Korotin, V. I. Anisimov, J. A. McLeod, E. Z. Kurmaev, A. Moewes, *Phys. Rev. B* **2010**, *81*, 035122
[77] Y. Deng, Y. Yu, Y. Song, J. Zhang, N. Z. Wang, Z. Sun, Y. Yi, Y. Z. Wu, S. Wu, J. Zhu, J. Wang, X. H. Chen, Y. Zhang, *Nature* **2018**, *563*, 94.
[78] Y. Shi, J. Kahn, B. Niu, Z. Fei, B. Sun, X. Cai, B. Francisco, D. Wu, Z. X. Shen, X. Xu, D. Cobden, Y. T. Cui, *Sci. Adv.* **2019**, *5* eaat8799.
[79] S.-Y. Xu, Q. Ma, H. Shen, V. Fatemi, S. Wu, T.-R. Chang, G. Chang, A. M. M. Valdivia, C.-K. Chan, Q. D. Gibson, J. Zhou, Z. Liu, K. Watanabe, T. Taniguchi, H. Lin, R. J. Cava, L. Fu, N. Gedik, P. Jarillo-Herrero, *Nat. Phys.* **2018**, *14*, 900.
[80] L. Liu, T. Zhai, *InfoMat* **2020**, *3*, 3.
[81] Z. Cai, B. Liu, X. Zou, H. M. Cheng, *Chem Rev* **2018**, *118*, 6091.
[82] X. Song, Z. Guo, Q. Zhang, P. Zhou, W. Bao, D. W. Zhang, *Small* **2017**, *13*, 1700098.
[83] X. Tong, K. Liu, M. Zeng, L. Fu, *InfoMat* **2019**, *1*, 460.
[84] A. Zavabeti, A. Jannat, L. Zhong, A. A. Haidry, Z. Yao, J. Z. Ou, *Nano-Micro Lett.* **2020**, *12*, 66.
[85] P. Wang, R. Selhorst, T. Emrick, A. Ramasubramaniam, M. D. Barnes, *J. Phys. Chem. C* **2018**, *123*, 1506.
[86] R. Selhorst, P. Wang, M. Barnes, T. Emrick, *Chem. Sci.* **2018**, *9*, 5047.

[87] R. C. Selhorst, E. Puodziukynaite, J. A. Dewey, P. Wang, M. D. Barnes, A. Ramasubramaniam, T. Emrick, *Chem. Sci.* **2016**, *7*, 4698.
[88] X. Tong, E. Ashalley, F. Lin, H. Li, Z. M. Wang, *Nano-Micro Lett.* **2015**, *7*, 203.
[89] J. Yuan, T. Sun, Z. Hu, W. Yu, K. Zhang, B. SuN, S. P. Lau, Q. Bao, S. Lin, S. Li, *ACS Appl Mater Interfaces* **2018**, *10*, 40614.
[90] N. Li, Q. Wang, C. Shen, Z. Wei, H. Yu, J. Zhao, X. Lu, G. Wang, C. He, L. Xie, J. Zhu, L. Du, R. Yang, D. Shi, G. Zhang, *Nat. Electron.* **2020**, *3*, 711.
[91] W. J. Yu, Y. Liu, H. Zhou, A. Yin, Z. Li, Y. Huang, X. Duan, *Nat. Nanotechnol.* **2013**, *8*, 952.
[92] J. Wang, Z. Li, H. Chen, G. Deng, X. Niu, *Nano-Micro Lett.* **2019**, *11*, 48.
[93] J. Pu, T. Takenobu, *Adv. Mater.* **2018**, *30*, 1707627.
[94] X. Gao, G. Bian, J. Zhu, *Journal of Materials Chemistry C* **2019**, *7*, 12835.
[95] J. Cheng, C. Wang, X. Zou, L. Liao, *Adv. Opt. Mater.* **2019**, *7*, 1800441.
[96] J. Shi, M. Hong, Z. Zhang, Q. Ji, Y. Zhang, *Coord. Chem. Rev.* **2018**, *376*, 1.
[97] T. Zhang, L. Fu, *Chem* **2018**, *4*, 671.
[98] X. Li, C. W. Magnuson, A. Venugopal, R. M. Tromp, J. B. Hannon, E. M. Vogel, L. Colombo, R. S. Ruoff, *Journal of the American Chemical Society* **2011**, *133*, 2816.
[99] O. J. Burton, F. C. Massabuau, V. P. Veigang-Radulescu, B. Brennan, A. J. Pollard, S. Hofmann, *ACS Nano* **2020**, *14*, 13593.
[100] J. H. Lee, E. K. Lee, W. J. Joo, Y. Jang, B. S. Kim, J. Y. Lim, S. H. Choi, S. J. Ahn, J. R. Ahn, M. H. Park, C. W. Yang, B. L. Choi, S. W. Hwang, D. Whang, *Science* **2014**, *344*, 286.
[101] W. A. de Heer, C. Berger, M. Ruan, M. Sprinkle, X. Li, Y. Hu, B. Zhang, J. Hankinson, E. Conrad, *Proc. Natl. Acad. Sci. U S A* **2011**, *108*, 16900.
[102] M. Wu, Z. Zhang, X. Xu, Z. Zhang, Y. Duan, J. Dong, R. Qiao, S. You, L. Wang, J. Qi, D. Zou, N. Shang, Y. Yang, H. Li, L. Zhu, J. Sun, H. Yu, P. Gao, X. Bai, Y. Jiang, Z. J. Wang, F. Ding, D. Yu, E. Wang, K. Liu, *Nature* **2020**, *581*, 406.
[103] Z. Q. Xu, Y. Zhang, S. Liu, C. Zheng, Y. L. Zhong, X. Xia, Z. Li, P. J. Sophia, M. Fuhrer, Y. B. Cheng, Q. Bao, *ACS Nano* **2015**, *9*, 6178.
[104] P. Liu, T. Luo, J. Xing, H. Xu, H. Hao, H. Liu, J. Dong, *Nanoscale Res. Lett.* **2017**, *12*, 558.
[105] C. Lee, Y. Hugen, B. Louis, T. Heinz, J. Hone, R. Sunmin, *ACS Nano* **2010**, *4*, 2695.
[106] J. Wu, H. Li, Z. Yin, H. Li, J. Liu, X. Cao, Q. Zhang, H. Zhang, *Small* **2013**, *9*, 3314.
[107] M. Velicky, G. E. Donnelly, W. R. Hendren, S. McFarland, D. Scullion, W. J. I. DeBenedetti, G. C. Correa, Y. Han, A. J. Wain, M. A. Hines, D. A. Muller, K. S. Novoselov, H. D. Abruna, R. M. Bowman, E. J. G. Santos, F. Huang, *ACS Nano* **2018**, *12*, 10463.
[108] Y. Huang, Y. H. Pan, R. Yang, L. H. Bao, L. Meng, H. L. Luo, Y. Q. Cai, G. D. Liu, W. J. Zhao, Z. Zhou, L. M. Wu, Z. L. Zhu, M. Huang, L. W. Liu, L. Liu, P. Cheng, K. H. Wu, S. B. Tian, C. Z. Gu, Y. G. Shi, Y. F. Guo, Z. G. Cheng, J. P. Hu, L. Zhao, G. H. Yang, E. Sutter, P. Sutter, Y. L. Wang, W. Ji, X. J. Zhou, H. J. Gao, *Nat. Commun.* **2020**, *11*, 2453.
[109] J. N. Coleman, M. Lotya, A. O'Neill, S. D. Bergin, P. J. King, U. Khan, K. Young, A. Gaucher, S. De, R. J. Smith, I. V. Shvets, S. K. Arora, G. Stanton, H. Y. Kim, K. Lee, G. T. Kim, G. S. Duesberg, T. Hallam, J. J. Boland, J. J. Wang, J. F. Donegan, J. C. Grunlan, G. Moriarty, A. Shmeliov, R. J. Nicholls, J. M. Perkins, E. M. Grieveson, K. Theuwissen, D. W. McComb, P. D. Nellist, V. Nicolosi, *Science* **2011**, *331*, 568.
[110] C. Xing, X. Chen, W. Huang, Y. Song, J. Li, S. Chen, Y. Zhou, B. Dong, D. Fan, X. Zhu, H. Zhang, *ACS Photonics* **2018**, *5*, 5055.
[111] P. Yang, X. Zou, Z. Zhang, M. Hong, J. Shi, S. Chen, J. Shu, L. Zhao, S. Jiang, X. Zhou, Y. Huan, C. Xie, P. Gao, Q. Chen, Q. Zhang, Z. Liu, Y. Zhang, *Nat. Commun.* **2018**, *9*, 979.
[112] M. Seol, M. H. Lee, H. Kim, K. W. Shin, Y. Cho, I. Jeon, M. Jeong, H. I. Lee, J. Park, H. J. Shin, *Adv. Mater.* **2020**, *32*, 2003542.

[113] K. Kang, S. Xie, L. Huang, Y. Han, P. Y. Huang, K. F. Mak, C. J. Kim, D. Muller, J. Park, *Nature* **2015**, *520*, 656.
[114] J. Wang, T. Li, Q. Wang, W. Wang, R. Shi, N. Wang, A. Amini, C. Cheng, *Mater. Today. Adv.* **2020**, *8*, 100098.
[115] P. Wang, S. Song, A. Najafi, C. Huai, P. Zhang, Y. Hou, S. Huang, H. Zeng, *ACS Nano* **2020**, *14*, 7370.
[116] H. Kim, D. Ovchinnikov, D. Deiana, D. Unuchek, A. Kis, *Nano Lett.* **2017**, *17*, 5056.
[117] J. Cai, X. Han, X. Wang, X. Meng, *Matter* **2020**, *2*, 587.
[118] Z. Yang, J. Hao, *Journal of Materials Chemistry C* **2016**, *4*, 8859.
[119] M. J. Chhowalla, Debdeep Zhang, Hua *Nat. Rev. Mater.* **2016**, *1*, 1.
[120] Y. Yoon, K. Ganapathi, S. Salahuddin, *Nano Lett.* **2011**, *11*, 3768.
[121] B. Radisavljevic, A. Radenovic, J. Brivio, V. Giacometti, A. Kis, *Nat. Nanotechnol.* **2011**, *6*, 147.
[122] C. Lan, Z. Zhou, Z. Zhou, C. Li, L. Shu, L. Shen, D. Li, R. Dong, S. Yip, J. C. Ho, *Nano Res.* **2018**, *11*, 3371.
[123] Y. Kim, H. Bark, G. H. Ryu, Z. Lee, C. Lee, *J. Phys. Condens. Matter* **2016**, *28*, 184002.
[124] H. Yu, M. Liao, W. Zhao, G. Liu, X. J. Zhou, Z. Wei, X. Xu, K. Liu, Z. Hu, K. Deng, S. Zhou, J. A. Shi, L. Gu, C. Shen, T. Zhang, L. Du, L. Xie, J. Zhu, W. Chen, R. Yang, D. Shi, G. Zhang, *ACS Nano* **2017**, *11*, 12001.
[125] X. Xu, S. Chen, S. Liu, X. Cheng, W. Xu, P. Li, Y. Wan, S. Yang, W. Gong, K. Yuan, P. Gao, Y. Ye, L. Dai, *J. Am. Chem. Soc.* **2019**, *141*, 2128.
[126] J. J. Pyeon, I. H. Baek, W. C. Lim, K. H. Chae, S. H. Han, G. Y. Lee, S. H. Baek, J. S. Kim, J. W. Choi, T. M. Chung, J. H. Han, C. Y. Kang, S. K. Kim, *Nanoscale* **2018**, *10*, 17712.
[127] H.-U. Kim, M. Kim, Y. Jin, Y. Hyeon, K. S. Kim, B.-S. An, C.-W. Yang, V. Kanade, J.-Y. Moon, G. Y. Yeom, D. Whang, J.-H. Lee, T. Kim, *Appl. Surf. Sci.* **2019**, *470*, 129.
[128] X. Ling, Y. H. Lee, Y. Lin, W. Fang, L. Yu, M. S. Dresselhaus, J. Kong, *Nano Lett.* **2014**, *14*, 464.
[129] S. S. Han, T. J. Ko, C. Yoo, M. S. Shawkat, H. Li, B. K. Kim, W. K. Hong, T. S. Bae, H. S. Chung, K. H. Oh, Y. Jung, *Nano Lett.* **2020**, *20*, 3925.
[130] K. K. Liu, W. Zhang, Y. H. Lee, Y. C. Lin, M. T. Chang, C. Y. Su, C. S. Chang, H. Li, Y. Shi, H. Zhang, C. S. Lai, L. J. Li, *Nano Lett.* **2012**, *12*, 1538.
[131] Y. R. Lim, W. Song, J. K. Han, Y. B. Lee, S. J. Kim, S. Myung, S. S. Lee, K. S. An, C. J. Choi, J. Lim, *Adv. Mater.* **2016**, *28*, 5025.
[132] S. Seo, H. Choi, S.-Y. Kim, J. Lee, K. Kim, S. Yoon, B. H. Lee, S. Lee, *Adv. Mater. Interfaces* **2018**, *5*, 1800524.
[133] C. R. Serrao, A. M. Diamond, S.-L. Hsu, L. You, S. Gadgil, J. Clarkson, C. Carraro, R. Maboudian, C. Hu, S. Salahuddin, *Appl. Phys. Lett.* **2015**, *106*, 052101.
[134] Q. He, P. Li, Z. Wu, B. Yuan, Z. Luo, W. Yang, J. Liu, G. Cao, W. Zhang, Y. Shen, P. Zhang, S. Liu, G. Shao, Z. Yao, *Adv. Mater.* **2019**, *31*, 1901578.
[135] Z. Chen, H. Liu, X. Chen, G. Chu, S. Chu, H. Zhang, *ACS Appl Mater Interfaces* **2016**, *8*, 20267.
[136] Z. Lin, Y. Zhao, C. Zhou, R. Zhong, X. Wang, Y. H. Tsang, Y. Chai, *Sci. Rep.* **2015**, *5*, 18596.
[137] A. George, C. Neumann, D. Kaiser, R. Mupparapu, T. Lehnert, U. Hübner, Z. Tang, A. Winter, U. Kaiser, I. Staude, A. Turchanin, *J. Phys.: Mater.* **2019**, *2*, 016001.
[138] R. Luo, W. W. Xu, Y. Zhang, Z. Wang, X. Wang, Y. Gao, P. Liu, M. Chen, *Nat. Commun.* **2020**, *11*, 1011.
[139] J.-G. Song, J. Park, W. C. Lee, Taejin, H. Jung, C. W. H. Lee, Sung-Hwan , J. M. J. Myoung, Jae-Hoon Kim, Soo-Hyun Lansalot-Matras, Clement , S.-H. Kim, C. Lansalot-Matras, H. Kim, *ACS Nano* **2013**, *7*, 11333.

[140] L. K. Tan, B. Liu, J. H. Teng, S. Guo, H. Y. Low, H. R. Tan, C. Y. Chong, R. B. Yang, K. P. Loh, *Nanoscale* **2014**, *6*, 10584.
[141] J. G. Eberhart, *J. Phys. Chem.* **1967**, *71*, 4125.
[142] P. Miskiewicz, S. Kotarba, J. Jung, T. Marszalek, M. Mas-Torrent, E. Gomar-Nadal, D. B. Amabilino, C. Rovira, J. Veciana, W. Maniukiewicz, J. Ulanski, *J. Appl. Phys.* **2008**, *104*, 054509.
[143] X. Liu, I. Balla, H. Bergeron, G. P. Campbell, M. J. Bedzyk, M. C. Hersam, *ACS Nano* **2016**, *10*, 1067.
[144] Y. Shi, W. Zhou, A. Y. Lu, W. Fang, Y. H. Lee, A. L. Hsu, S. M. Kim, K. K. Kim, H. Y. Yang, L. J. Li, J. C. Idrobo, J. Kong, *Nano Lett.* **2012**, *12*, 2784.
[145] S. V. Mandyam, H. M. Kim, M. Drndić, *J. Phys.: Mater.* **2020**, *3*, 024008.
[146] H. Tang, H. Zhang, X. Chen, Y. Wang, X. Zhang, P. Cai, W. Bao, *Sci. China. Inf. Sci.* **2019**, *62*, 220401.
[147] A. T. Hoang, K. Qu, X. Chen, J. H. Ahn, *Nanoscale* **2021**, *13*, 615.
[148] L. Tang, T. Li, Y. Luo, S. Feng, Z. Cai, H. Zhang, B. Liu, H. M. Cheng, *ACS Nano* **2020**, *14*, 4646.
[149] J. Zhou, J. Lin, X. Huang, Y. Zhou, Y. Chen, J. Xia, H. Wang, Y. Xie, H. Yu, J. Lei, D. Wu, F. Liu, Q. Fu, Q. Zeng, C. H. Hsu, C. Yang, L. Lu, T. Yu, Z. Shen, H. Lin, B. I. Yakobson, Q. Liu, K. Suenaga, G. Liu, Z. Liu, *Nature* **2018**, *556*, 355.
[150] S. Y. Kim, J. Kwak, C. V. Ciobanu, S. Y. Kwon, *Adv. Mater.* **2019**, *31*, 1804939.
[151] S. Grimme, J. Antony, S. Ehrlich, H. Krieg, *J. Chem. Phys.* **2010**, *132*, 154104.
[152] M. Bosi, *RSC Adv.* **2015**, *5*, 75500.
[153] P. Wang, J. Lei, J. Qu, S. Cao, H. Jiang, M. He, H. Shi, X. Sun, B. Gao, W. Liu, *Chem. Mater.* **2019**, *31*, 873.
[154] Y. Cai, K. Xu, W. Zhu, *Mater. Res. Express* **2018**, *5*, 095904.
[155] S. H. Choi, Y. J. Kim, W. Yang, K. K. Kim, *J. Korean. Phys. Soc.* **2019**, *74*, 1032.
[156] A. Kozhakhmetov, R. Torsi, C. Y. Chen, J. A. Robinson, *J. Phys.: Mater.* **2020**, *4*, 012001.
[157] A. Singh, M. Moun, M. Sharma, A. Barman, A. Kumar Kapoor, R. Singh, *Appl. Surf. Sci.* **2021**, *538*, 148201.
[158] R. Ionescu, B. Campbell, R. Wu, E. Aytan, A. Patalano, I. Ruiz, S. W. Howell, A. E. McDonald, T. E. Beechem, K. A. Mkhoyan, M. Ozkan, C. S. Ozkan, *Sci Rep* **2017**, *7*, 6419.
[159] O. A. Abbas, I. Zeimpekis, H. Wang, A. H. Lewis, N. P. Sessions, M. Ebert, N. Aspiotis, C. C. Huang, D. Hewak, S. Mailis, P. Sazio, *Sci. Rep.* **2020**, *10*, 1696.
[160] C. Muratore, A. A. Voevodin, N. R. Glavin, *Thin Solid Films* **2019**, *688*.
[161] T. Alam, B. Wang, R. Pulavarthy, M. A. Haque, C. Muratore, N. Glavin, A. K. Roy, A. A. Voevodin, *Appl. Phys. Lett.* **2014**, *105*, 213110.
[162] H. Heo, J. H. Sung, J.-H. Ahn, F. Ghahari, T. Taniguchi, K. Watanabe, P. Kim, M.-H. Jo, *Adv. Electron. Mater.* **2017**, *3*, 1600375.
[163] J. Zhang, D. J. Xue, X. Zhan, Z. Li, D. Zeng, H. Song, *ACS Appl Mater Interfaces* **2017**, *9*, 27102.
[164] H. Q. Ta, D. J. Perello, D. L. Duong, G. H. Han, S. Gorantla, V. L. Nguyen, A. Bachmatiuk, S. V. Rotkin, Y. H. Lee, M. H. Rummeli, *Nano Lett.* **2016**, *16*, 6403.
[165] J. D. Yao, Z. Q. Zheng, G. W. Yang, *Prog. Mater Sci.* **2019**, *106*, 100573.
[166] L. El Bouanani, M. I. Serna, M. N. Hasan S, B. L. Murillo, S. Nam, H. Choi, H. N. Alshareef, M. A. Quevedo-Lopez, *ACS Appl Mater Interfaces* **2020**, *12*, 51645.
[167] F. Godel, V. Zatko, C. Carrétéro, A. Sander, M. Galbiati, A. Vecchiola, P. Brus, O. Bezencenet, B. Servet, M.-B. Martin, B. Dlubak, P. Seneor, *ACS Appl. Nano Mater.* **2020**, *3*, 7908.
[168] M. M. Juvaid, M. S. Ramachandra Rao, *Materials Today: Proceedings* **2021**, *35*, 494.

[169] M. Yan, E. Wang, X. Zhou, G. Zhang, H. Zhang, K. Zhang, W. Yao, N. Lu, S. Yang, S. Wu, T. Yoshikawa, K. Miyamoto, T. Okuda, Y. Wu, P. Yu, W. Duan, S. Zhou, *2D Mater.* **2017**, *4*, 045015.
[170] F. Presel, H. Tetlow, L. Bignardi, P. Lacovig, C. A. Tache, S. Lizzit, L. Kantorovich, A. Baraldi, *Nanoscale* **2018**, *10*, 7396.
[171] S. Liu, X. Yuan, Y. Zou, Y. Sheng, C. Huang, E. Zhang, J. Ling, Y. Liu, W. Wang, C. Zhang, J. Zou, K. Wang, F. Xiu, *npj 2D Mater. and Appl.* **2017**, *1*, 30.
[172] J. Zhang, H. Yu, W. Chen, X. Tian, D. Liu, M. Cheng, G. Xie, W. Yang, R. Yang, X. Bai, D. Shi, G. Zhang, *ACS Nano* **2014**, *8*, 6024.
[173] J. You, M. D. Hossain, Z. Luo, *Nano Converg* **2018**, *5*, 26.
[174] J. Park, K.-H. Xue, M. Mouis, F. Triozon, A. Cresti, *Phys. Rev. B* **2019**, *100*, 235403.
[175] H.-P. Komsa, A. V. Krasheninnikov, *Adv. Electron. Mater.* **2017**, *3*, 1600468.
[176] R. Dong, I. Kuljanishvili, *J Vac Sci Technol B Nanotechnol Microelectron* **2017**, *35*, 030803.
[177] L. Chen, B. Liu, M. Ge, Y. Ma, A. N. Abbas, C. Zhou, *ACS Nano* **2015**, *9*, 8368.
[178] Z. Wang, H. Yang, S. Zhang, J. Wang, K. Cao, Y. Lu, W. Hou, S. Guo, X. A. Zhang, L. Wang, *Nanoscale* **2019**, *11*, 22440.
[179] M. C. Chang, P. H. Ho, M. F. Tseng, F. Y. Lin, C. H. Hou, I. K. Lin, H. Wang, P. P. Huang, C. H. Chiang, Y. C. Yang, I. T. Wang, H. Y. Du, C. Y. Wen, J. J. Shyue, C. W. Chen, K. H. Chen, P. W. Chiu, L. C. Chen, *Nat. Commun.* **2020**, *11*, 3682.
[180] S. Zhou, L. Gan, D. Wang, H. Li, T. Zhai, *Nano Res.* **2018**, *11*, 2909.
[181] F. Zhang, K. Momeni, M. A. AlSaud, A. Azizi, M. F. Hainey, J. M. Redwing, L.-Q. Chen, N. Alem, *2D Mater.* **2017**, *4*, 025029.
[182] B. Pan, K. Zhang, C. Ding, Z. Wu, Q. Fan, T. Luo, L. Zhang, C. Zou, S. Huang, *ACS Appl Mater Interfaces* **2020**, *12*, 35337.
[183] F. Lan, R. Yang, Y. Xu, S. Qian, S. Zhang, H. Cheng, Y. Zhang, *Nanomaterials (Basel)* **2018**, *8*.
[184] S. Hu, X. Wang, L. Meng, X. Yan, *J. Mater. Sci.* **2017**, *52*, 7215.
[185] Y. Li, K. Zhang, F. Wang, Y. Feng, Y. Li, Y. Han, D. Tang, B. Zhang, *ACS Appl Mater Interfaces* **2017**, *9*, 36009.
[186] H. Zhou, C. Wang, J. C. Shaw, R. Cheng, Y. Chen, X. Huang, Y. Liu, N. O. Weiss, Z. Lin, Y. Huang, X. Duan, *Nano Lett.* **2015**, *15*, 709.
[187] Y. F. Lim, K. Priyadarshi, F. Bussolotti, P. K. Gogoi, X. Cui, M. Yang, J. Pan, S. W. Tong, S. Wang, S. J. Pennycook, K. E. J. Goh, A. T. S. Wee, S. L. Wong, D. Chi, *ACS Nano* **2018**, *12*, 1339.
[188] J. D. Cain, E. D. Hanson, V. P. Dravid, *J. Appl. Phys.* **2018**, *123*, 204304.
[189] L. Yang, C. Xie, J. Jin, R. N. Ali, C. Feng, P. Liu, B. Xiang, *Nanomaterials (Basel)* **2018**, *8*, 463.
[190] P. Yang, A.-G. Yang, L. Chen, J. Chen, Y. Zhang, H. Wang, L. Hu, R.-J. Zhang, R. Liu, X.-P. Qu, Z.-J. Qiu, C. Cong, *Nano Res.* **2019**, *12*, 823.
[191] S. Wang, Y. Rong, Y. Fan, M. Pacios, H. Bhaskaran, K. He, J. H. Warner, *Chem. Mater.* **2014**, *26*, 6371.
[192] Y. Rong, Y. Fan, A. Leen Koh, A. W. Robertson, K. He, S. Wang, H. Tan, R. Sinclair, J. H. Warner, *Nanoscale* **2014**, *6*, 12096.
[193] R. Yue, Y. Nie, L. A. Walsh, R. Addou, C. Liang, N. Lu, A. T. Barton, H. Zhu, Z. Che, D. Barrera, L. Cheng, P.-R. Cha, Y. Chabal, J. W. P. Hsu, J. Kim, M. J. Kim, L. Colombo, R. M. Wallace, K. Cho, C. L. Hinkle, *2D Mater.* **2017**, *4*, 045019.
[194] J. Chen, X. Zhao, S. J. Tan, H. Xu, B. Wu, B. Liu, D. Fu, W. Fu, D. Geng, Y. Liu, W. Liu, W. Tang, L. Li, W. Zhou, T. C. Sum, K. P. Loh, *J. Am. Chem. Soc.* **2017**, *139*, 1073.
[195] W. Chen, J. Zhao, J. Zhang, L. Gu, Z. Yang, X. Li, H. Yu, X. Zhu, R. Yang, D. Shi, X. Lin, J. Guo, X. Bai, G. Zhang, *J. Am. Chem. Soc.* **2015**, *137*, 15632.
[196] S. Zhao, L. Wang, L. Fu, *iScience* **2019**, *20*, 527.

[197] J. Dong, L. Zhang, X. Dai, F. Ding, *Nat. Commun.* **2020**, *11*, 5862.
[198] A. Aljarb, Z. Cao, H. L. Tang, J. K. Huang, M. Li, W. Hu, L. Cavallo, L. J. Li, *ACS Nano* **2017**, *11*, 9215.
[199] S. K. Pandey, R. Das, P. Mahadevan, *ACS Omega* **2020**, *5*, 15169.
[200] Y. Sun, D. Wang, Z. Shuai, *J. Phys. Chem. C* **2016**, *120*, 21866.
[201] D. Wickramaratne, F. Zahid, R. K. Lake, *J. Chem. Phys.* **2014**, *140*, 124710.
[202] R. A. B. Villaos, C. P. Crisostomo, Z.-Q. Huang, S.-M. Huang, A. A. B. Padama, M. A. Albao, H. Lin, F.-C. Chuang, *npj 2D Mater. and Appl.* **2019**, *3*, 2.
[203] S. L. Shang, G. Lindwall, Y. Wang, J. M. Redwing, T. Anderson, Z. K. Liu, *Nano Lett.* **2016**, *16*, 5742.
[204] Y. Kim, J. G. Song, Y. J. Park, G. H. Ryu, S. J. Lee, J. S. Kim, P. J. Jeon, C. W. Lee, W. J. Woo, T. Choi, H. Jung, H. B. Lee, J. M. Myoung, S. Im, Z. Lee, J. H. Ahn, J. Park, H. Kim, *Sci. Rep.* **2016**, *6*, 18754.
[205] X. Zhang, S. Y. Teng, A. C. M. Loy, B. S. How, W. D. Leong, X. Tao, *Nanomaterials (Basel)* **2020**, *10*, 1012.
[206] Z. Cai, Y. Lai, S. Zhao, R. Zhang, J. Tan, S. Feng, J. Zou, L. Tang, J. Lin, B. Liu, H.-M. Cheng, *Natl. Sci. Rev.* **2020**, *0*, 1.
[207] J. Yu, X. Hu, H. Li, X. Zhou, T. Zhai, *Journal of Materials Chemistry C* **2018**, *6*, 4627.
[208] L. Tang, J. Tan, H. Nong, B. Liu, H.-M. Cheng, *Acc. Mater. Res.* **2020**, *2*, 36.
[209] T. Park, M. Leem, H. Lee, W. Ahn, H. Kim, J. Kim, E. Lee, Y.-H. Kim, H. Kim, *J. Phys. Chem. C* **2017**, *121*, 27693.
[210] M. O'Brien, N. McEvoy, T. Hallam, H. Y. Kim, N. C. Berner, D. Hanlon, K. Lee, J. N. Coleman, G. S. Duesberg, *Sci. Rep.* **2014**, *4*, 7374.
[211] J. Chen, K. Shao, W. Yang, W. Tang, J. Zhou, Q. He, Y. Wu, C. Zhang, X. Li, X. Yang, Z. Wu, J. Kang, *ACS Appl Mater Interfaces* **2019**, *11*, 19381.
[212] S. S. Withanage, H. Kalita, H. S. Chung, T. Roy, Y. Jung, S. I. Khondaker, *ACS Omega* **2018**, *3*, 18943.
[213] X. Song, W. Zan, H. Xu, S. Ding, P. Zhou, W. Bao, D. W. Zhang, *2D Mater.* **2017**, *4*, 025051.
[214] P. Taheri, J. Wang, H. Xing, J. F. Destino, M. M. Arik, C. Zhao, K. Kang, B. Blizzard, L. Zhang, P. Zhao, S. Huang, S. Yang, F. V. Bright, J. Cerne, H. Zeng, *Mater. Res. Express* **2016**, *3*, 075009.
[215] C. R. Wu, X. R. Chang, C. H. Wu, S. Y. Lin, *Sci. Rep.* **2017**, *7*, 42146.
[216] M. E. McConney, N. R. Glavin, A. T. Juhl, M. H. Check, M. F. Durstock, A. A. Voevodin, T. E. Shelton, J. E. Bultman, J. Hu, M. L. Jespersen, M. K. Gupta, R. D. Naguy, J. G. Colborn, A. Haque, P. T. Hagerty, R. E. Stevenson, C. Muratore, *J. Mater. Res.* **2016**, *31*, 967.
[217] J. Tao, J. Chai, X. Lu, L. M. Wong, T. I. Wong, J. Pan, Q. Xiong, D. Chi, S. Wang, *Nanoscale* **2015**, *7*, 2497.
[218] J. G. Song, J. Park, W. Lee, T. Choi, H. Jung, C. W. Lee, S. H. Hwang, J. M. Myoung, J. H. Jung, S. H. Kim, C. Lansalot-Matras, H. Kim, *ACS Nano* **2013**, *7*, 11333.
[219] Y. Huang, L. Liu, *Sci. China Mater.* **2019**, *62*, 913.
[220] S. B. Basuvalingam, M. A. Bloodgood, M. A. Verheijen, W. M. M. Kessels, A. A. Bol, *ACS Appl. Nano Mater.* **2021**, *4*, 514.
[221] H. Cun, M. Macha, H. Kim, K. Liu, Y. Zhao, T. LaGrange, A. Kis, A. Radenovic, *Nano Res.* **2019**, *12*, 2646.
[222] T. Kim, J. Mun, H. Park, D. Joung, M. Diware, C. Won, J. Park, S. H. Jeong, S. W. Kang, *Nanotechnology* **2017**, *28*, 18LT01.
[223] M. Chubarov, T. H. Choudhury, D. R. Hickey, S. Bachu, T. Zhang, A. Sebastian, A. Bansal, H. Zhu, N. Trainor, S. Das, M. Terrones, N. Alem, J. M. Redwing, *ACS Nano* **2021**, *15*, 2532.
[224] P. Wang, S. Luo, L. Boyle, H. Zeng, S. Huang, *Nanoscale* **2019**, *11*, 17065.

[225] S. H. Choi, B. Stephen, J. H. Park, J. S. Lee, S. M. Kim, W. Yang, K. K. Kim, *Sci. Rep.* **2017**, *7*, 1983.
[226] P. Yang, Z. Zhang, M. Sun, F. Lin, T. Cheng, J. Shi, C. Xie, Y. Shi, S. Jiang, Y. Huan, P. Liu, F. Ding, C. Xiong, D. Xie, Y. Zhang, *ACS Nano* **2019**, *13*, 3649.
[227] J. Jeon, S. K. Jang, S. M. Jeon, G. Yoo, Y. H. Jang, J. H. Park, S. Lee, *Nanoscale* **2015**, *7*, 1688.
[228] Y.-T. Ho, C.-H. Ma, T.-T. Luong, L.-L. Wei, T.-C. Yen, W.-T. Hsu, W.-H. Chang, Y.-C. Chu, Y.-Y. Tu, K. P. Pande, E. Y. Chang, *Phys. Status Solidi RRL* **2015**, *9*, 187.
[229] J. D. Cain, F. Shi, J. Wu, V. P. Dravid, *ACS Nano* **2016**, *10*, 5440.
[230] L. Wu, W. Yang, G. Wang, *npj 2D Mater. and Appl.* **2019**, *3*, 6.
[231] M. S. Sokolikova, C. Mattevi, *Chem. Soc. Rev.* **2020**, *49*, 3952.
[232] Z. Qian, L. Jiao, L. Xie, *Chinese Journal of Chemistry* **2020**, *38*, 753.
[233] S. Park, C. Kim, S. O. Park, N. K. Oh, U. Kim, J. Lee, J. Seo, Y. Yang, H. Y. Lim, S. K. Kwak, G. Kim, H. Park, *Adv. Mater.* **2020**, *32*, 2001889.
[234] L. Yang, W. Zhang, J. Li, S. Cheng, Z. Xie, H. Chang, *ACS Nano* **2017**, *11*, 1964.
[235] X.-J. Yan, Y.-Y. Lv, L. Li, X. Li, S.-H. Yao, Y.-B. Chen, X.-P. Liu, H. Lu, M.-H. Lu, Y.-F. Chen, *npj Quantum Mater.* **2017**, *2*, 31.
[236] Y. Xiao, M. Zhou, J. Liu, J. Xu, L. Fu, *Sci. China Mater.* **2019**, *62*, 759.
[237] T. Lehnert, M. Ghorbani-Asl, J. Köster, Z. Lee, A. V. Krasheninnikov, U. Kaiser, *ACS Appl. Nano Mater.* **2019**, *2*, 3262.
[238] Y. C. Lin, D. O. Dumcenco, Y. S. Huang, K. Suenaga, *Nat. Nanotechnol.* **2014**, *9*, 391.
[239] J. Zhu, Z. Wang, H. Yu, N. Li, J. Zhang, J. Meng, M. Liao, J. Zhao, X. Lu, L. Du, R. Yang, D. Shi, Y. Jiang, G. Zhang, *J. Am. Chem. Soc.* **2017**, *139*, 10216.
[240] S. Song, D. H. Keum, S. Cho, D. Perello, Y. Kim, Y. H. Lee, *Nano Lett.* **2016**, *16*, 188.
[241] R. Kappera, D. Voiry, S. E. Yalcin, B. Branch, G. Gupta, A. D. Mohite, M. Chhowalla, *Nat. Mater.* **2014**, *13*, 1128.
[242] M. Naz, T. Hallam, N. C. Berner, N. McEvoy, R. Gatensby, J. B. McManus, Z. Akhter, G. S. Duesberg, *ACS Appl Mater Interfaces* **2016**, *8*, 31442.
[243] T. A. Empante, Y. Zhou, V. Klee, A. E. Nguyen, I. H. Lu, M. D. Valentin, S. A. Naghibi Alvillar, E. Preciado, A. J. Berges, C. S. Merida, M. Gomez, S. Bobek, M. Isarraraz, E. J. Reed, L. Bartels, *ACS Nano* **2017**, *11*, 900.
[244] J. Kim, Z. Lee, *Appl. Microsc.* **2018**, *48*, 43.
[245] H. U. Kim, V. Kanade, M. Kim, K. S. Kim, B. S. An, H. Seok, H. Yoo, L. E. Chaney, S. I. Kim, C. W. Yang, G. Y. Yeom, D. Whang, J. H. Lee, T. Kim, *Small* **2020**, *16*, 1905000.
[246] R. Wang, Q. Shao, Q. Yuan, P. Sun, R. Nie, X. Wang, *Appl. Surf. Sci.* **2020**, *504*, 144320.
[247] T. A. Loh, D. H. Chua, A. T. Wee, *Sci. Rep.* **2015**, *5*, 18116.
[248] M. Hafeez, L. Gan, H. Li, Y. Ma, T. Zhai, *Adv. Funct. Mater.* **2016**, *26*, 4551.
[249] W. Choi, N. Choudhary, G. H. Han, J. Park, D. Akinwande, Y. H. Lee, *Mater. Today* **2017**, *20*, 116.
[250] I. Katsouras, D. Zhao, M. J. Spijkman, M. Li, P. W. Blom, D. M. de Leeuw, K. Asadi, *Sci. Rep.* **2015**, *5*, 12094.
[251] V. Dobrovolsky, F. Sizov, *J. Appl. Phys.* **2012**, *112*, 124517.
[252] E. Ko, J. Shin, C. Shin, *Nano Converg* **2018**, *5*, 2.
[253] H. Xu, H. Zhang, Y. Liu, S. Zhang, Y. Sun, Z. Guo, Y. Sheng, X. Wang, C. Luo, X. Wu, J. Wang, W. Hu, Z. Xu, Q. Sun, P. Zhou, J. Shi, Z. Sun, D. W. Zhang, W. Bao, *Adv. Funct. Mater.* **2019**, *29*, 1805614.
[254] S. Song, Y. Sim, S.-Y. Kim, J. H. Kim, I. Oh, W. Na, D. H. Lee, J. Wang, S. Yan, Y. Liu, J. Kwak, J.-H. Chen, H. Cheong, J.-W. Yoo, Z. Lee, S.-Y. Kwon, *Nat. Electron.* **2020**, *3*, 207.
[255] T. Liu, C. Han, D. Xiang, K. Han, A. Ariando, W. Chen, *Adv. Sci.* **2020**, *7*, 2002393.

[256] A. T. Hoang, A. K. Katiyar, H. Shin, N. Mishra, S. Forti, C. Coletti, J. H. Ahn, *ACS Appl Mater Interfaces* **2020**, *12*, 44335.
[257] S. Riazimehr, A. Bablich, D. Schneider, S. Kataria, V. Passi, C. Yim, G. S. Duesberg, M. C. Lemme, *Solid-State Electronics* **2016**, *115*, 207.
[258] X. An, F. Liu, Y. J. Jung, S. Kar, *Nano Lett.* **2013**, *13*, 909.
[259] X. Li, M. Zhu, M. Du, Z. Lv, L. Zhang, Y. Li, Y. Yang, T. Yang, X. Li, K. Wang, H. Zhu, Y. Fang, *Small* **2016**, *12*, 595.
[260] Y. Zhao, J. Qiao, Z. Yu, P. Yu, K. Xu, S. P. Lau, W. Zhou, Z. Liu, X. Wang, W. Ji, Y. Chai, *Adv. Mater.* **2017**, *29*, 1604230.
[261] Z. X. Zhang, Z. Long-Hui, X. W. Tong, Y. Gao, C. Xie, Y. H. Tsang, L. B. Luo, Y. C. Wu, *J. Phys. Chem. Lett.* **2018**, *9*, 1185.
[262] L.-H. Zeng, S.-H. Lin, Z.-J. Li, Z.-X. Zhang, T.-F. Zhang, C. Xie, C.-H. Mak, Y. Chai, S. P. Lau, L.-B. Luo, Y. H. Tsang, *Adv. Funct. Mater.* **2018**, *28*, 1705970.
[263] P. Lv, X. Zhang, X. Zhang, W. Deng, J. Jie, *IEEE Electron Device Lett.* **2013**, *34*, 1337.
[264] L.-H. Zeng, D. Wu, S.-H. Lin, C. Xie, H.-Y. Yuan, W. Lu, S. P. Lau, Y. Chai, L.-B. Luo, Z.-J. Li, Y. H. Tsang, *Adv. Funct. Mater.* **2019**, *29*, 1806878.
[265] J. Sławińska, F. T. Cerasoli, H. Wang, S. Postorino, A. Supka, S. Curtarolo, M. Fornari, M. Buongiorno Nardelli, *2D Mater.* **2019**, *6*, 025012.
[266] C. K. Safeer, J. Ingla-Aynes, F. Herling, J. H. Garcia, M. Vila, N. Ontoso, M. R. Calvo, S. Roche, L. E. Hueso, F. Casanova, *Nano Lett.* **2019**, *19*, 1074.
[267] X. Wen, Z. Gong, D. Li, *InfoMat* **2019**, *1*, 317.
[268] Z. Wu, B. T. Zhou, X. Cai, P. Cheung, G. B. Liu, M. Huang, J. Lin, T. Han, L. An, Y. Wang, S. Xu, G. Long, C. Cheng, K. T. Law, F. Zhang, N. Wang, *Nat. Commun.* **2019**, *10*, 611.
[269] T. H. Choudhury, X. Zhang, Z. Y. Al Balushi, M. Chubarov, J. M. Redwing, *Annu. Rev. Mater. Res.* **2020**, *50*, 155.
[270] L. Peters, C. Ó Coileáin, P. Dluzynski, R. Siris, G. S. Duesberg, N. McEvoy, *Phys. Status Solidi (a)* **2020**, *217*, 2000073.
[271] C. Ahn, J. Lee, H. U. Kim, H. Bark, M. Jeon, G. H. Ryu, Z. Lee, G. Y. Yeom, K. Kim, J. Jung, Y. Kim, C. Lee, T. Kim, *Adv. Mater.* **2015**, *27*, 5223.
[272] Y. Zhao, J. G. Song, G. H. Ryu, K. Y. Ko, W. J. Woo, Y. Kim, D. Kim, J. H. Lim, S. Lee, Z. Lee, J. Park, H. Kim, *Nanoscale* **2018**, *10*, 9338.
[273] T. Wu, X. Zhang, Q. Yuan, J. Xue, G. Lu, Z. Liu, H. Wang, H. Wang, F. Ding, Q. Yu, X. Xie, M. Jiang, *Nat. Mater.* **2016**, *15*, 43.
[274] J. Tan, S. Li, B. Liu, H.-M. Cheng, *Small Structures* **2020**, *2*, 2000093.
[275] B. Tang, B. Che, M. Xu, Z. P. Ang, J. Di, H.-J. Gao, H. Yang, J. Zhou, Z. Liu, *Small Structures* **2021**, *1*, 2000153.